\documentclass{jfm}

\usepackage{graphicx}
\usepackage{subfigure}
\usepackage{newtxtext}
\usepackage{newtxmath}
\usepackage{natbib}
\usepackage{hyperref}
\usepackage{booktabs}
\usepackage{threeparttable}
\usepackage{makecell}
\usepackage{amsmath}
\hypersetup{
    colorlinks = true,
    urlcolor   = blue,
    citecolor  = black,
}

\newcommand{\RomanNumeralCaps}[1]
\title{Disambigutaion Decomposition of Mean Skin Friction and Heat Flux on Arbitrary-Curvature Surface}

\author{Ming-zhi Tang\aff{1,2}
  Wen-feng Zhou\aff{1,2}
  Yan-chao Hu\aff{1,2}
  \corresp{\email{huyanchao@cardc.cn}},
  Gang Wang\aff{1,2}
  Ming Fang\aff{1,2}
 \and Yan-guang Yang\aff{2,3}
 \corresp{\email{yangyanguang@cardc.cn}}}

\affiliation{\aff{1}Hypervelocity Aerodynamics Institute, China Aerodynamics Research and Development Center, Mianyang 621000, China
\aff{2}Laboratory of Aerodynamics in Multiple Flow Regimes (LAMFR), China Aerodynamics Research and Development Center, Mianyang 621000, China
\aff{3}China Aerodynamics Research and Development Center, Mianyang 621000, China}

\begin{document}
\maketitle

\begin{abstract}
Since it is difficult to apply the existing method of friction and heat flux decomposition on the complex surface, a combined decomposition method of friction and heat flux with clear physical interpretation is proposed, which is based on FIK and RD decomposition method and can be applied to arbitrary surface. Based on this method, the aerothermodynamic characteristics of bistable states of curved compression ramps are analyzed from the perspective of energy transformation.  The results show that the decrease of friction in the interaction region of the attachment state and the minimum values of friction in the separation bubble are all caused by the energy injection of the work by the adverse pressure gradient. The peak friction is mainly induced by the viscous dissipation, and its position is affected by the mechanical energy transport. The peak heat flux is mainly induced by viscous dissipation, and the enthalpy transport of the separation state plays a greater role in the peak heat flux generation than that of the attachment state. These results indicate that reducing viscous dissipation is a potential way for realizing friction and heat flux control simultaneously.
\end{abstract}

\begin{keywords}

\end{keywords}

%{\bf MSC Codes }  {\it(Optional)} Please enter your MSC Codes here

\section{Introduction}
\label{sec:intro}
\par Deflection of control surface such as body flaps, elevons and rudders may cause intense shock wave/boundary layer interaction (SBLI). SBLI may yield significant flow separation and lead to significant decrease in control effectiveness and the excessive increase of the heat flux, and even lose control of aircraft (\cite{simeonides1994experimental,simeonides1995experimental,babinsky2014shock,zhang2020hypersonic}). During this process, multistable states of the flow may be encountered, which can lead to multistable systems manifested as the dependence of a system on its evolutionary history, and ubiquitous in aerospace flow systems (\cite{hu2020bistable}). Multistable states can lead to multiple values of aerodynamic coefficients such as lift and drag (\cite{yang2008an,mccroskey1982unsteady,mueller1985the,biber1993hysteresis,mittal2000prediction}). The bistable states of the regular and Mach reflection is found when changing the incident angle of the shock (\cite{hornung1979transition,chpoun1995reconsideration,vuillon1995reconsideration,chpoun1995numerical,ivanov2001flow}), and the mechanism is explained by Hu et al. with minimal dissipation theory (\cite{hu2021mechanism}). Recently, Hu et al observed bistable states of separation/attachment in curved compression ramp (CCR) flows induced by variation of attack angle via numerical simulation (\cite{hu2020bistable}) and offer mathematical demonstration of the existence of bistable states in CCR flows (\cite{hu2021existence}). Mechanism of separation hysteresis and the aerothermodynamics of bistable states in CCR flows are discussed by Zhou et al (\cite{zhou2021mechanism}) and Tang et al (\cite{tang2021aerothermodynamic}).

Wall friction and heat flux are the results of velocity and temperature redistribution in the boundary layer. And it has always been the goal of researchers to accurately and quantitatively understand this redistribution process, which is of great significance in basic research and engineering applications. Friction and heat flux decomposition provides the possibility to achieve this goal.  

Currently, there are mainly two decomposition methods commonly used, namely, friction decomposition method based on momentum equation integration proposed by Fukagata et al. (\cite{fukagata2002contribution}) (denoted as FIK decomposition), and decomposition method based on Galileo transformation and kinetic energy equation proposed by Renard and Deck (\cite{renard2016theoretical}) (denoted as RD decomposition). These two methods are widely used to analyze the mechanism of friction generation and reduction by flow control. Recently, some researchers extended these two decomposition ideas to analyze heat flux generation, such as Zhang Peng (\cite{zhang2020contribution} \cite{zhang2022exact}), Sun Dong (\cite{sun2021decomposition}). However, these analyses mainly focused on the equilibrium flow on one-dimensional surface such as channel flow and flat plate flow, and could not analyze the flow characteristics on arbitrary-curvature surface. In addition, there's still no consensus on the physical interpretation of the two decomposition methods. Therefore, the number of integral orders in FIK decomposition and the selection of reference velocity in RD decomposition are arbitrary, which makes it difficult to accurately quantify the friction and heat flow generation mechanism and the effect of flow control in experiments or engineering.  

Therefore, it is necessary to obtain a decomposition method with clear physical interpretation and suitable for any profile, which is helpful for accurate and effective evaluation of friction and heat flow control methods. Based on the power-energy transformation, a physically clear form of friction and heat flow decomposition applicable to arbitrary-curvature surface is given below. The aerothermodynamic characteristics of curved compression ramp (CCR) flows with bistable states are analyzed, and the mechanism of friction and heat flux generation, as well as the Reynolds analogy is deeply analyzed.

% * <ajohns@cambridge.org> 2018-06-14T10:05:40.418Z:
%
% ^.

\section{Decomposition Method}\label{sec:rules_submission}
 
Renard and Deck analyzed the work-energy transformation process of differential form in their paper (equation (2.5) in Ref.(\cite{renard2016theoretical}). This paper will discuss this process from the perspective of integration, and demonstrate the uniqueness of transformation velocity $u_b$, in the following.  

\subsection{Derivation of mean skin friction decomposition}
\begin{figure*} %[htbp]
	\centering
%	 Use the relevant command to insert your figure file.
%	 For example, with the graphicx package use
	\includegraphics[width = 0.8\columnwidth]{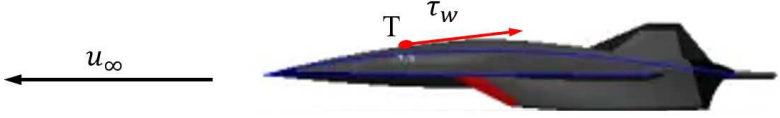}
	\caption{Schematic diagram of vehicle flight.}
	\label{fig:vech_cf}       % Give a unique label
\end{figure*}

\noindent The following derivation is based on the rotated equations of the coordinate system. Following the process of RD decomposition, we can obtain the intermediate result as  
\begin{align}\label{eq:step1}
\tilde{u}_{b} \bar{\tau}_{w}=&\int_{0}^{\infty} \bar{\tau}_{y x} \frac{\partial \tilde{u}}{\partial y} d y+\int_{0}^{\infty} \bar{\rho}\left(-\overline{u^{\prime \prime} v^{\prime \prime}}\right) \frac{\partial \tilde{u}}{\partial y} d y+\int_{0}^{\infty}\left(\tilde{u}-\tilde{u}_{b}\right)\left[\bar{\rho}\left(\tilde{u} \frac{\partial \tilde{u}}{\partial x}+\tilde{v} \frac{\partial \tilde{u}}{\partial y}\right)\right] d y \\\nonumber
&+ \int_{0}^{\infty}\left(\tilde{u}-\tilde{u}_{b}\right) \frac{\partial}{\partial x}\left(\bar{\rho} \widetilde{u^{\prime \prime} u^{\prime \prime}}-\bar{\tau}_{x x}\right) d y+\int_{0}^{\infty}\left(\tilde{u}-\tilde{u}_{b}\right) \frac{\partial \bar{P}}{\partial x} d y
\end{align}
In the relative coordinate system where the aircraft flies while the flow is still, the left side of equation \ref{eq:step1} represents the aircraft wall transferring work to the fluid through friction, and the input work can also be expressed as  

\begin{align}
	\mathrm{SW}=\vec{\tau} \cdot \vec{u}_{\infty}=\bar{\tau}_{w} u_{\infty} \cos \varphi
\end{align}

where $\varphi$ is the included Angle between the local wall surface and $u_{\infty}$. In fact, this work is equal to the energy required for the vehicle to overcome the drag (point T in figure .\ref{fig:vech_cf}). According to the analysis of equation \ref{eq:step1}, $u_b$ is unique as

\begin{align}\label{eq:step2}
	u_b=u_{\infty} \cos \varphi
\end{align}
rather than arbitrary.

Combining equation \ref{eq:step1} and \ref{eq:step2}, the final form of friction decomposition can be obtained
\begin{align}
\underbrace{C_{f}}_{C_{f, 0}}=&\underbrace{\frac{2}{\rho_{\infty} u_{\infty}^{3} \cos \varphi} \int_{0}^{\infty} \bar{\tau}_{y x} \frac{\partial \tilde{u}}{\partial y} d y}_{C_{f, L}}+\underbrace{\frac{2}{\rho_{\infty} u_{\infty}^{3}} \int_{0}^{\infty} \bar{\rho}\left(-\overline{u^{\prime \prime} v^{\prime \prime}}\right) \frac{\partial \tilde{u}}{\partial y} d y}_{C_{f, T}} \\\nonumber
&+\underbrace{\frac{2}{\rho_{\infty} u_{\infty}^{3} \cos \varphi} \int_{0}^{\infty}\left(\tilde{u}-u_{\infty} \cos \varphi\right)\left[\bar{\rho}\left(\tilde{u} \frac{\partial \tilde{u}}{\partial x}+\tilde{v} \frac{\partial \tilde{u}}{\partial y}\right)\right] d y}_{C_{f, M T}} \\\nonumber
&+\underbrace{\frac{2}{\rho_{\infty} u_{\infty}^{3} \cos \varphi} \int_{0}^{\infty}\left(\tilde{u}-u_{\infty} \cos \varphi\right) \frac{\partial}{\partial x}\left(\bar{\rho} \widetilde{u}^{\prime \prime} u^{\prime \prime}-\bar{\tau}_{x x}\right) d y}_{C_{f, s}} \\\nonumber
&+\underbrace{\frac{2}{\rho_{\infty} u_{\infty}^{3} \cos \varphi} \int_{0}^{\infty}\left(\tilde{u}-u_{\infty} \cos \varphi\right) \frac{\partial \bar{P}}{\partial x} d y}_{C_{f, P}}
\end{align}
In above derivations, a possible puzzle is that $\varphi$ is a local quantity, therefore the Galilean transformation and the rotation of coordinate may be different at various positions. In fact, this problem does not affect the decomposition method, since the local transformation is sufficient to analyze the local friction characteristics, compared with the uniform transformation for the whole flow field.   

\subsection{Derivation of mean heat flux decomposition}

The following derivation is consistent with that of Zhang et al., but we will propose an explicit physical depiction and physical interpretation. The equation for average static enthalpy is 
\begin{align}\label{eq:step1_ch}
\frac{\partial \overline{\rho e}}{\partial t}+\frac{\partial \overline{\rho h u_{j}}}{\partial x_{j}}=\frac{\partial \overline{q_{j}}}{\partial x_{j}}+\overline{\tau_{i, j} \frac{\partial u_{i}}{\partial x_{j}}}+\overline{u_{j} \frac{\partial p}{\partial x_{j}}}
\end{align}
where $e=c_v T$ is the internal energy, the static enthalpy $h=e+p / \rho = c_p T$, and the heat flux $\overline{q_{j}}=\overline{\kappa \partial T / \partial x_{j}}$.

For steady flows, i.e. $\partial / \partial t=0$, integrating the energy equation from 0 (the wall) to y along the normal direction and the no-slip condition on the wall is considered as $u_w = v_w =0$, we can obtain

\begin{align}\label{eq:step2_ch}
	q_{w}=q_{y}-\left.\overline{\rho h v}\right|_{y}+\int_{0}^{y} \overline{\tau_{i, j} \frac{\partial u_{i}}{\partial x_{j}}} d y_{1}+\int_{0}^{y} u_{j} \frac{\partial p}{\partial x_{j}} d y_{1}+\int_{0}^{y}-\frac{\partial \overline{\rho h u}}{\partial x} d y_{1}  
\end{align}

\begin{figure*}
	\centering
%	 Use the relevant command to insert your figure file.
%	 For example, with the graphicx package use
	\includegraphics[width = 0.6\columnwidth]{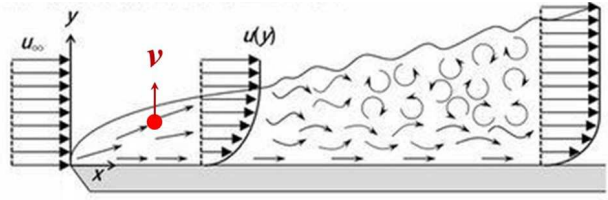}
%	 figure caption is below the figure
	\caption{Normalwise enthalpy transfer.}
	\label{fig:cht_ch}       % Give a unique label
\end{figure*}

The terms at the right hand side of equation. \ref{eq:step2_ch} are energy conduction at y, enthalpy transfer along the normal direction (red arrow in figure \ref{fig:cht_ch}), viscous dissipation from the wall to y, the work by pressure gradient, and enthalpy transport along the streamwise direction, respectively. This equation expresses the local equilibrium relation on the wall, but also gives the essence of heat flux generation on the wall, namely, the redistribution of temperature (enthalpy) in the boundary layer along the normal direction. This process includes the local generation of enthalpy in the boundary layer, and the transport along the streamwise and the normal direction. Among them, the convective heat transport is the most obvious way to enhance the redistribution of boundary layer temperature along the normal direction. And the total normally convected enthalpy from the outer boundary layer to the wall (or in reverse) can be obtained by integrating this term from 0 to $\delta$, with $\delta$ the boundary layer thickness.

Similar to FIK decomposition and Zhang et al. 's derivation, integrating equation \ref{eq:step2_ch} by parts from 0 to $\delta$ along the normal direction, we can get the final form of mean heat flux decomposition

\begin{align}\label{eq:ch_dec}
C_{h}=&\underbrace{\frac{1}{\rho_{\infty} u_{\infty}\left(h_{a w}-h_{w}\right) \delta} \int_{0}^{\delta} q_{y} d y}_{C_{h}^{L}}+\underbrace{\frac{1}{\rho_{\infty} u_{\infty}\left(h_{a w}-h_{w}\right) \delta} \int_{0}^{\delta}-\overline{\rho h v} d y}_{C_{h}^{\text {CHT }}} \\\nonumber
&+\underbrace{\frac{1}{\rho_{\infty} u_{\infty}\left(h_{a w}-h_{w}\right) \delta} \int_{0}^{\delta}(\delta-y) \tau_{i, j} \frac{\partial u_{i}}{\partial x_{j}} d y}_{C_{h}^{\text {WMS }}} \\\nonumber
&+\underbrace{\frac{1}{\rho_{\infty} u_{\infty}\left(h_{a w}-h_{w}\right) \delta} \int_{0}^{\delta}(\delta-y) u_{j} \frac{\partial p}{\partial x_{j}}}_{C_{h}^{WP}} d y  \\\nonumber
&+\underbrace{\frac{1}{\rho_{\infty} u_{\infty}\left(h_{a w}-h_{w}\right) \delta} \int_{0}^{\delta}(\delta-y)\left(-\frac{\partial \overline{\rho h u}}{\partial x}\right) d y}_{C_{h}^{X T}}
\end{align}

where the Stanton number $C_{h}=\frac{q_{w}}{\rho_{\infty} u_{\infty}\left(h_{a w}-h_{w}\right)}$. This relation shows that the heat flux at the wall can be decomposed into the heat transfer $C_{h}^{L}$, the convective heat transport $C_{h}^{\text {CHT }}$, the work from viscous stress $C_{h}^{\text {WMS }}$ and pressure gradient $C_{h}^{WP}$, as well as streamwise transport $C_{h}^{X T}$.

\subsection{Mesh-independent data conversion method for arbitrary-curvature surface}
\noindent According to the definition, the data processing of heat flow and friction decomposition should be integrated along the normal direction. The existing heat flow and friction decomposition analysis is usually established on one-dimensional surfaces such as flat plates and tube flows, and the integration along the normal direction can be directly carried out along the grid lines. And transformation of the flow field is in need for arbitrary-curvature surface. The transformation methods include coordinate scaling, coordinate rotation, or directly to deal with the equations in curved surface coordinates.  

\begin{itemize}
	\item Coordinate scaling can be carried out by Jacobi transformation, i.e. transform flow field of any shape into a square. Curved wall will transform to the flow direction, and integration along grid lines is feasible as the grid lines have been transformed to the normal potion.
	
	
	\item If the equations in curved coordinate system are dealt with directly, the equations are clear(equation \ref{eq:cur_equ}), but  these equations are usually related to the local flow direction, which is difficult to carry out in practical operation.
	
	\begin{align}\label{eq:cur_equ}
		\rho\left[\frac{u_{s}}{h} \frac{\partial u_{s}}{\partial s}+u_{n} \frac{\partial u_{s}}{\partial n}+\frac{\kappa u_{s} u_{n}}{h}\right]=-\frac{1}{h} \frac{d p}{d x}+\frac{\partial \tau}{\partial n}
	\end{align}	
	
	\item Rotation of the computational domain is also a feasible method. That is, the flow field is rotated to the flow direction of the concerned position without changing the essence of the inertial system (figure \ref{fig:rotate}). 
	
	\begin{figure*}  %[htbp]
		\centering
		%\includegraphics{rotate}
		\includegraphics[width = 1.0\columnwidth]{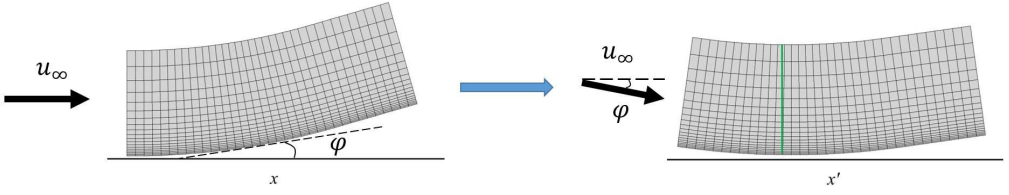}
		\caption{Coordinate rotating diagram.}
		\label{fig:rotate}       % Give a unique label
	\end{figure*}
	
\end{itemize}

The selection of integral path direction is discussed below. Except for the definition, although the information on grid lines can transform to the normal position of the surface, it represents the upstream or downstream information of the actual position, which is difficult to represent local real information. Besides, integral results will be associated with grid set and is not unique. Therefore, integration along the normal direction is necessary to obtain the real friction and heat decomposition results. Specifically, grid interpolation can be performed, that is, normal grid is set locally (green line in figure \ref{fig:rotate}) and interpolated in the computational domain to obtain flow parameters along the normal direction.  

\section {Simulation details and decomposition verification}
\subsection {Simulation details}
	Direct numerical simulations (DNS) of 2D CCR flow based on OpenCFD (\cite{li2010direct}) are carried out in the present research. The curved wall is an arc with the curvature radius $R=\frac{L}{\sin\frac{\phi}{2}}$, where $\phi$ is the turning angle and $L=25mm$ with $x=-L$ is the starting point of the curved wall. The unit Reynolds number $Re_{\infty}$ ($1 mm^{-1}$) of the flow is $3000$, $Pr = 0.7$, the specific heat ratio $\gamma = 1.4$. The flow conditions of inflow Mach number $Ma_{\infty}$, the normalized wall temperature $T_{w}$ ($T_{w} = \hat{T}_{w}/\hat{T}_{0}$, where $\hat{T}_{0}=108.1K$ is the inflow temperature) and the turning angle $\phi$ are summarised in table \ref{tab:flow_cond}. Other flow settings are consistent with Ref. (\cite{zhou2021mechanism} \cite{tang2021aerothermodynamic}) and numerical validations have been provided in Refs.(\cite{hu2020bistable,li2010direct,tang2021aerothermodynamic}).\\\indent

\begin{table}
	\begin{center}
		\def~{\hphantom{0}}
		\begin{tabular}{lccc}
			$Ma_{\infty}$  &  $T_{w}$  &  $\phi$ ($ ^\circ$) \\[3pt]
			5.0  &  1.5 &  \makecell[l]{16(Att), 17(Att/Sep),\\ 18(Att/Sep), 19(Sep)} \\   
			6.0  &  1.5  &  \makecell[l]{17(Att), 18(Att/Sep),\\ 19(Att/Sep), 20(Att/Sep),\\ 21(Att/Sep), 22(Att/Sep),\\ 23(Att/Sep), 24(Sep)} \\    
			6.0  &  2.0  &  \makecell[l]{16(Att), 17(Att/Sep),\\ 18(Att/Sep), 19(Sep)} \\ 
			6.0  &  \makecell[l]{1.25(Att),1.5(Att/Sep),\\1.75(Att/Sep),2.0(Att/Sep),\\2.25(Sep)} & 18
		\end{tabular}
		\caption{Flow conditions and CCR configurations for current simulations. (For each case, "Att" and "Sep" represent attachment and separation states, respectively.)}
		\label{tab:flow_cond}
	\end{center}
\end{table}

\subsection {Decomposition verification}

\begin{figure*} 
%	\hspace{-10mm}
	\subfigure[\label{subfig:cf_vf}]{\includegraphics[width = 0.5\columnwidth]{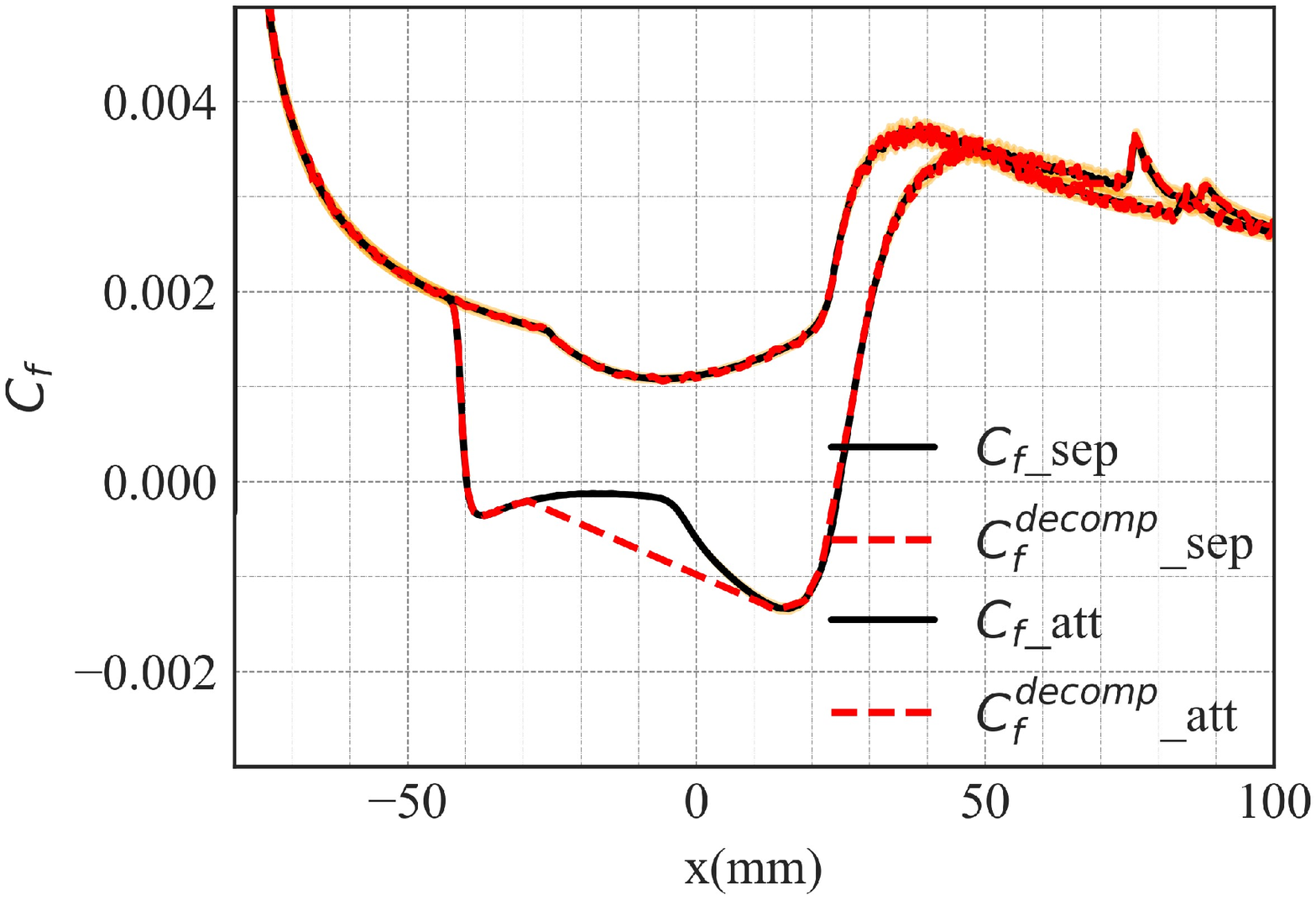}}
%	\hspace{-1.5mm}		
	\subfigure[\label{subfig:ch_vf}]{\includegraphics[width = 0.5\columnwidth]{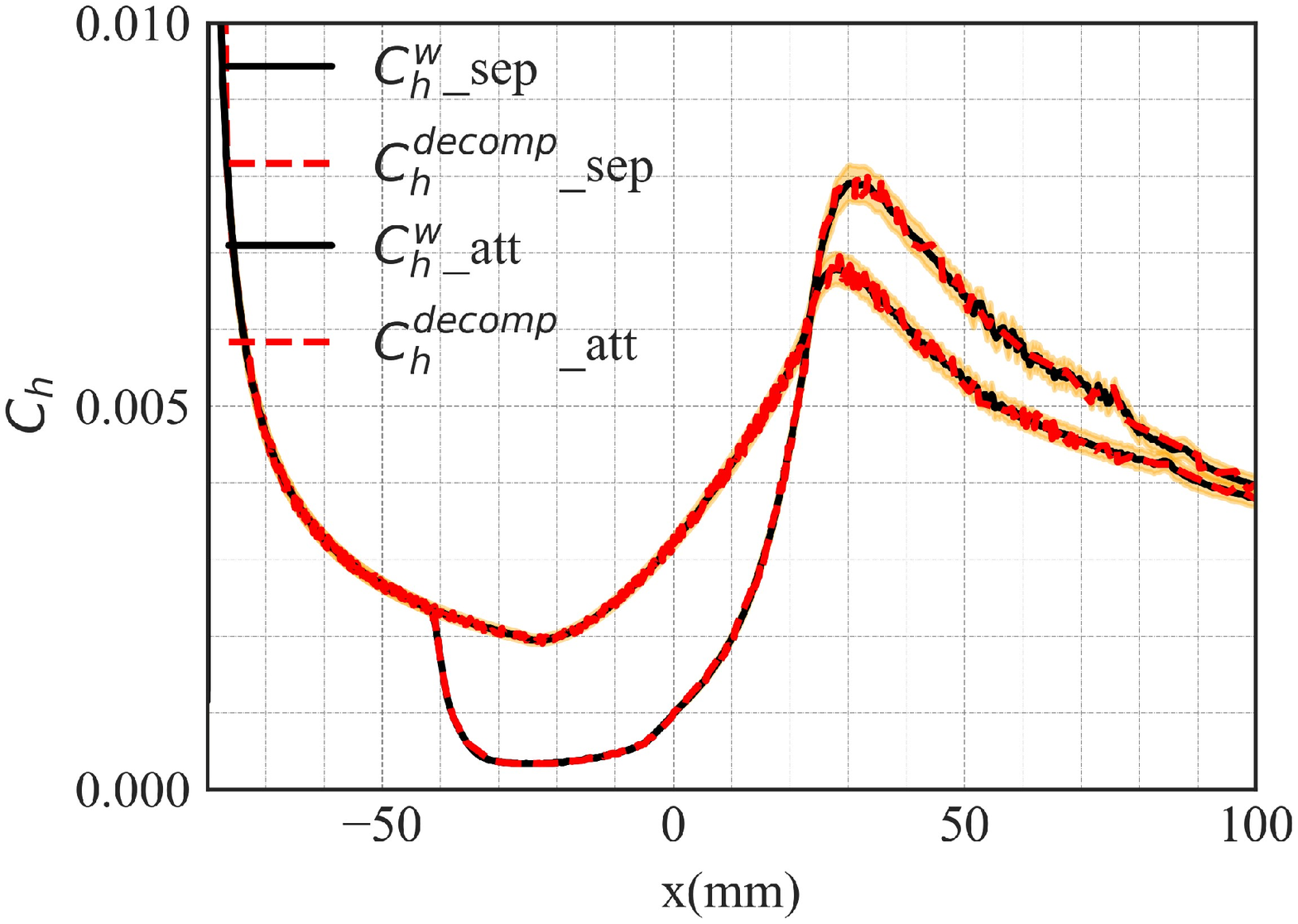}} 
%	\hspace{-10mm}
	\caption{Comparison of the results between the numerical simulation and the decomposition of friction(a) and heat flux(b) decomposition (the orange area is 3\% error band)} \label{fig:dec_vf}
\end{figure*}

Figure \ref{fig:dec_vf} compared the result between the numerical simulation and the decomposition. And the relative errors are within 3\%, indicating the effectiveness of the present decomposition methods, even they are applied to curved walls. Besides, the results also indicate the following analysis based on the present data is credible.

\section {Streamwise evolution of $C_{f}$}
\begin{figure*} 
%	\hspace{-1.5mm}
%	\centerline
	\subfigure[\label{subfig:cf_bal}]{\includegraphics[width = 0.49\columnwidth]{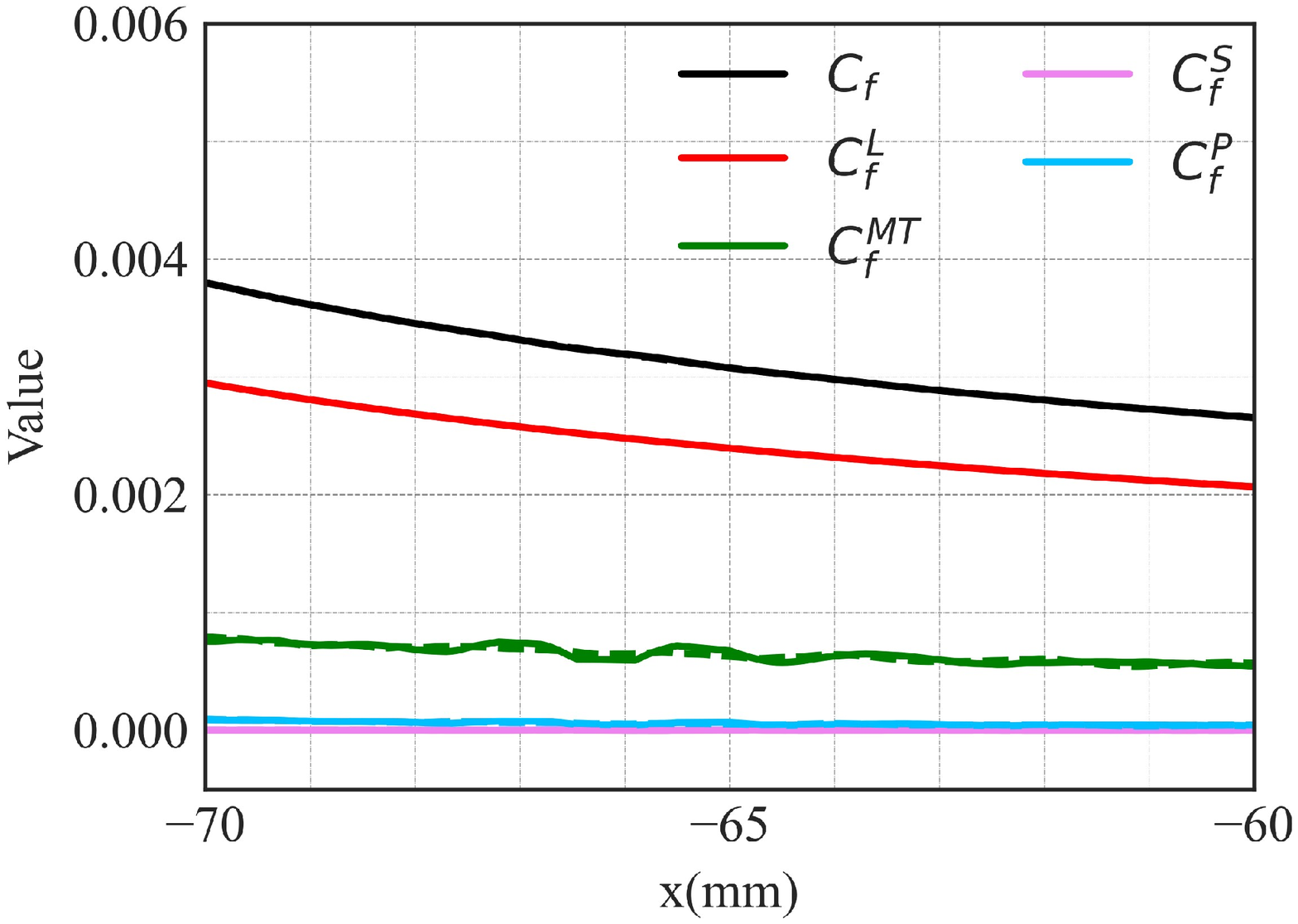}}
%	\hspace{-1.5mm}		
	\subfigure[\label{subfig:cf_sep}]{\includegraphics[width = 0.5\columnwidth]{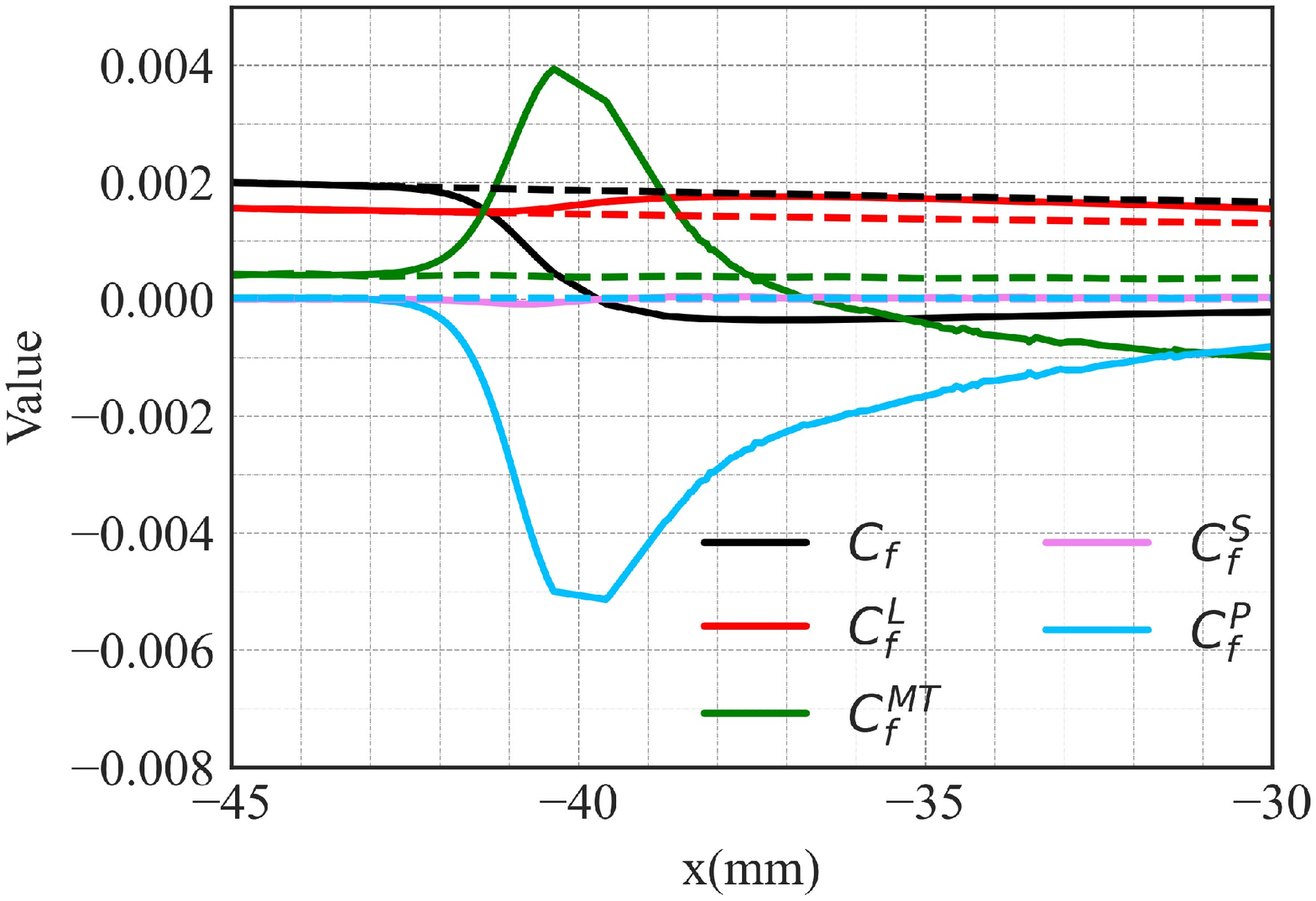}} 
%	\hspace{-1.5mm}
	\subfigure[\label{subfig:cf_reatt}]{\includegraphics[width = 0.5\columnwidth]{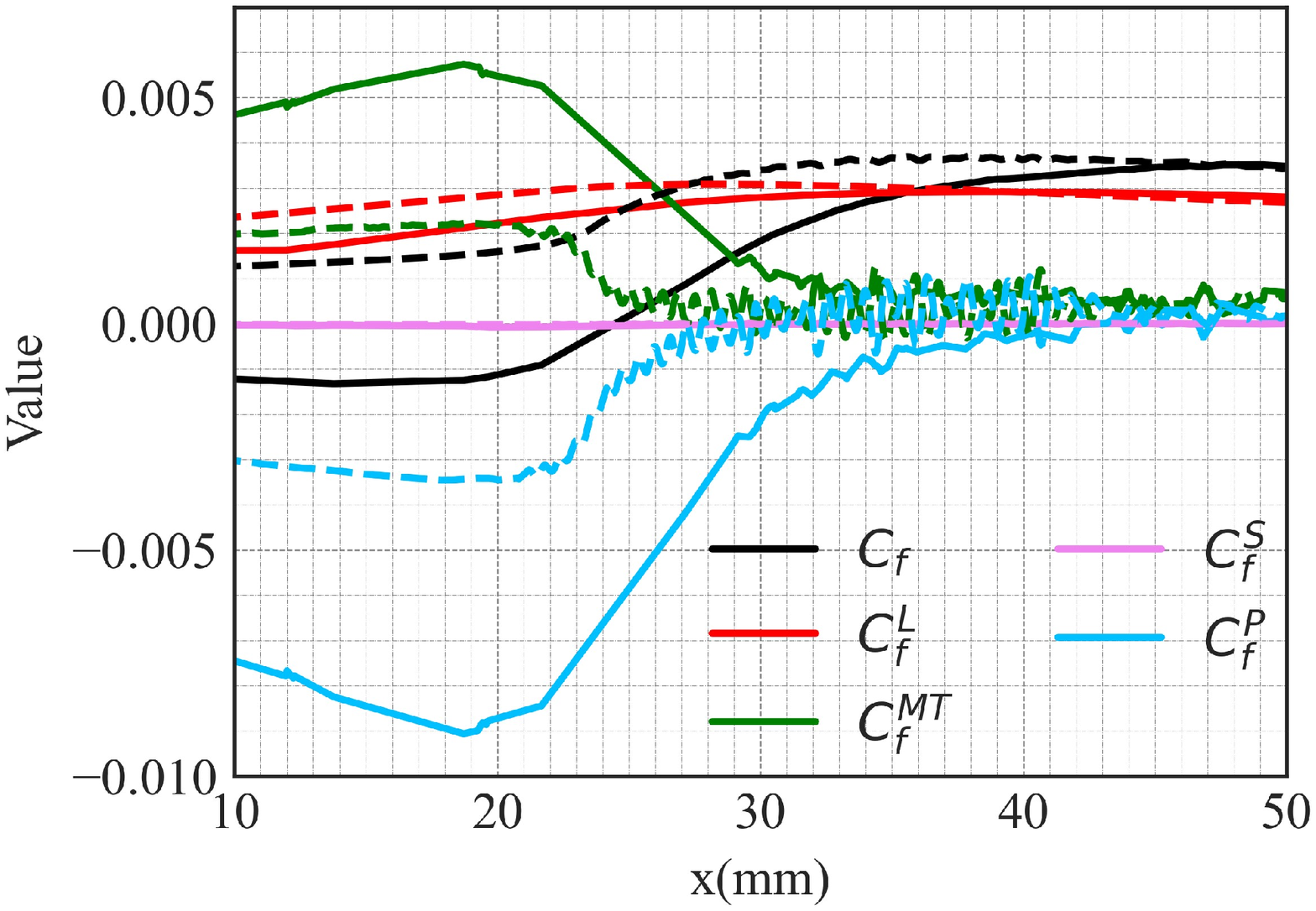}}
%	\hspace{-1.5mm}		
	\subfigure[\label{subfig:cf_recv}]{\includegraphics[width = 0.5\columnwidth]{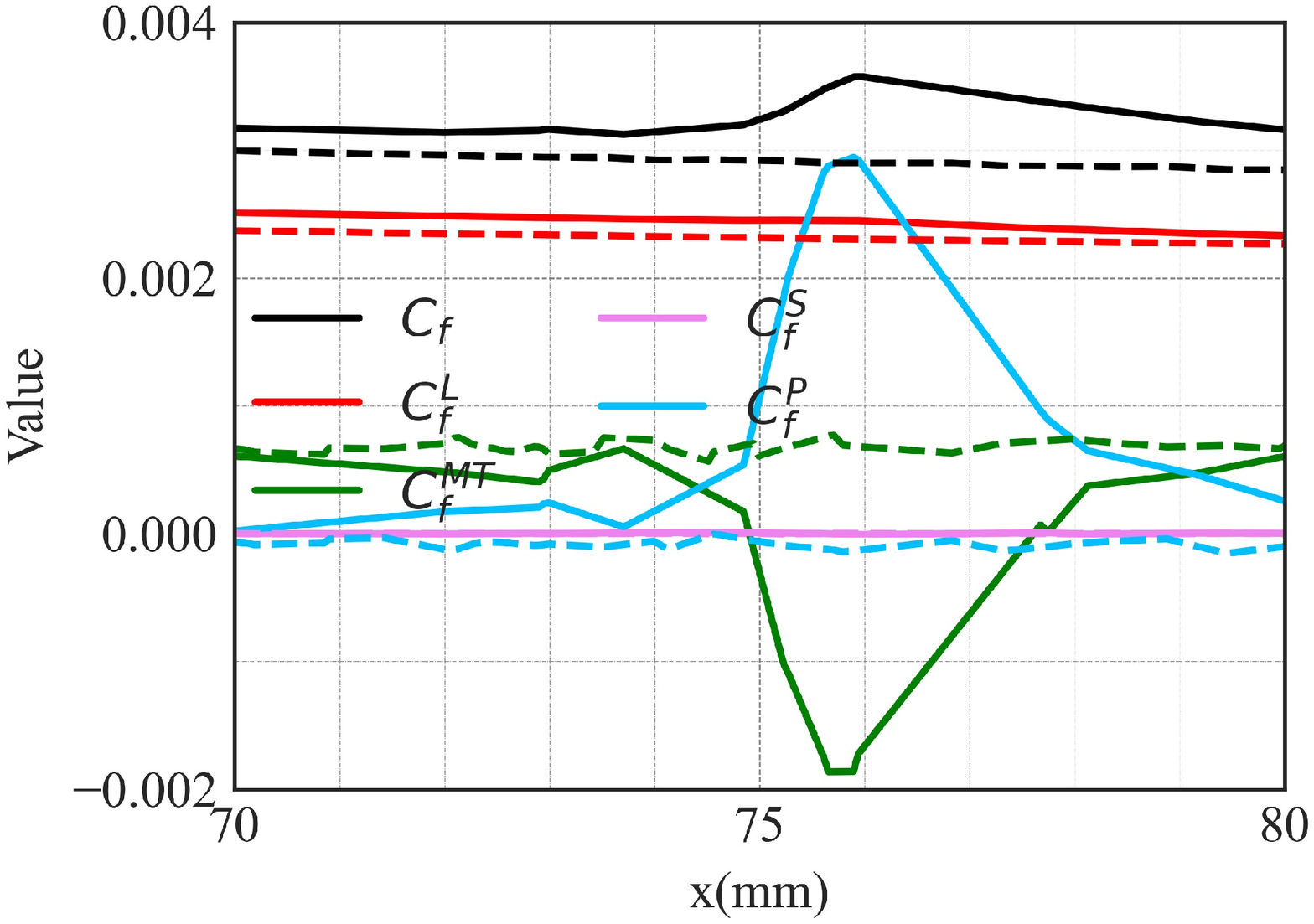}} 
%	\hspace{-1.5mm}	
	\caption{Comparison of friction decomposition of separated/attached states at different streamwise positions (the solid lines are separated states, and dashed lines are attached states.) (a) equilibrium flow region; (b) separation region; (c) reattachment and peak friction region; (d) recovery region.} \label{fig:cf_stream}
\end{figure*}

In this section, streamwise evolution of $C_{f}$ at different locations (including the equilibrium flow region, the separation region, the reattachment and peak friction region, as well as recovery boundary layer region) are analyzed, as shown in figure \ref{fig:cf_stream}. It should be noted that for the friction decomposition, the following analysis stands on the relative coordinate system by the Galilean transformation.

In the equilibrium flow region (figure \ref{subfig:cf_bal}), the flow of the separation state remains attached there, which behaves the same as the attachment state. And the dominant factor for friction generation is the work done by viscous stresses $C_{f}^{L}$, which also indicates that the work done by wall friction from the vehicle to the boundary layer is mainly transformed into viscous dissipation. The streamwise kinetic energy transport $C_{f}^{MT}$ plays the secondary role, which results from the boundary layer development along the streamwise direction, as the boundary layer thickens and the kinetic energy contained in the boundary layer increases. But the increase of kinetic energy also shows a gradual downward trend.

In the separation region of the separation state (figure \ref{subfig:cf_sep}), the attachment state maintains the characteristics of the equilibrium flow region, because the flow is not disturbed. The wall friction of the separation state decreases obviously, which is the combined effect of the work done by the adverse pressure gradient (APG) $C_{f}^{P}$ and $C_{f}^{MT}$. $C_{f}^{P}$ caused by the separation shock wave is negative, indicating that the work by APG is injected into the boundary layer. Here we should emphasize that the positive and negative values represent the direction from the work done by the wall friction to other terms, and the work from other terms to the wall friction, of energy transformation, respectively. And when we discuss the work done by APG, we refer to the absolute value in the following. The APG will inevitably change the boundary layer velocity profile, which is manifested as the obvious increase of kinetic energy of the boundary layer, but $C_{f}^{L}$ is still a major factor in the energy transformation process. The work done by APG leads to a significant decrease of the requirement for work by the wall friction for the maintenance of energy in the boundary layer, and even results in negative work by local wall friction, which represents the flow separation.

The reattachment region (figure \ref{subfig:cf_reatt}) shows similar results to the separation region, where $C_{f}^{P}$ caused by the reattachment shock waves plays the dominant role.

\begin{figure*}
	\centering
	%	 Use the relevant command to insert your figure file.
	%	 For example, with the graphicx package use
	\includegraphics[width = 0.5\columnwidth]{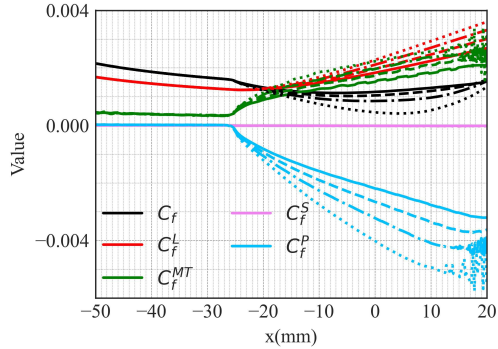}
	%	 figure caption is below the figure
	\caption{Comparison of friction decomposition with different wall temperature ratios (different linetypes correspond to different wall temperature ratios).}
	\label{fig:cf_tw}       % Give a unique label
\end{figure*}

$C_{f}^{P}$ is the coupling between APG and the boundary layer velocity profile. For the attachment state, with the increase of $\phi$ or the increase of $T_{w}$, $C_{f}$ shows a downward trend. Since the pressure rise on the surface under different wall temperatures ratio of the attachment states are nearly isentropic compression, the pressure gradient distribution is basically the same. Therefore we compare the influence of  wall temperature ratio to discuss the influence of the boundary layer velocity profile below, as shown in figure \ref{fig:cf_tw}. In the equilibrium flow region, variation of wall temperature ratio has no effect on the various terms of friction decomposition. While on the curved wall, with the increase of wall temperature ratio, $C_{f}^{L}$ and $C_{f}^{MT}$ increase, but $C_{f}^{P}$ decreases more. The results shows that larger velocity profile deficit (higher shape factor, as discussed by Zhou et al.\cite{zhou2021mechanism}) results in more $C_{f}^{P}$, even with similar APG. Ttherefore it is more difficult to dissipate the energy injection by APG, and the flow is more likely to separate.

\begin{figure*} 
	%	\hspace{-1.5mm}
%	\centerline
	\subfigure[\label{subfig:cf2_sep}]{\includegraphics[width = 0.5\columnwidth]{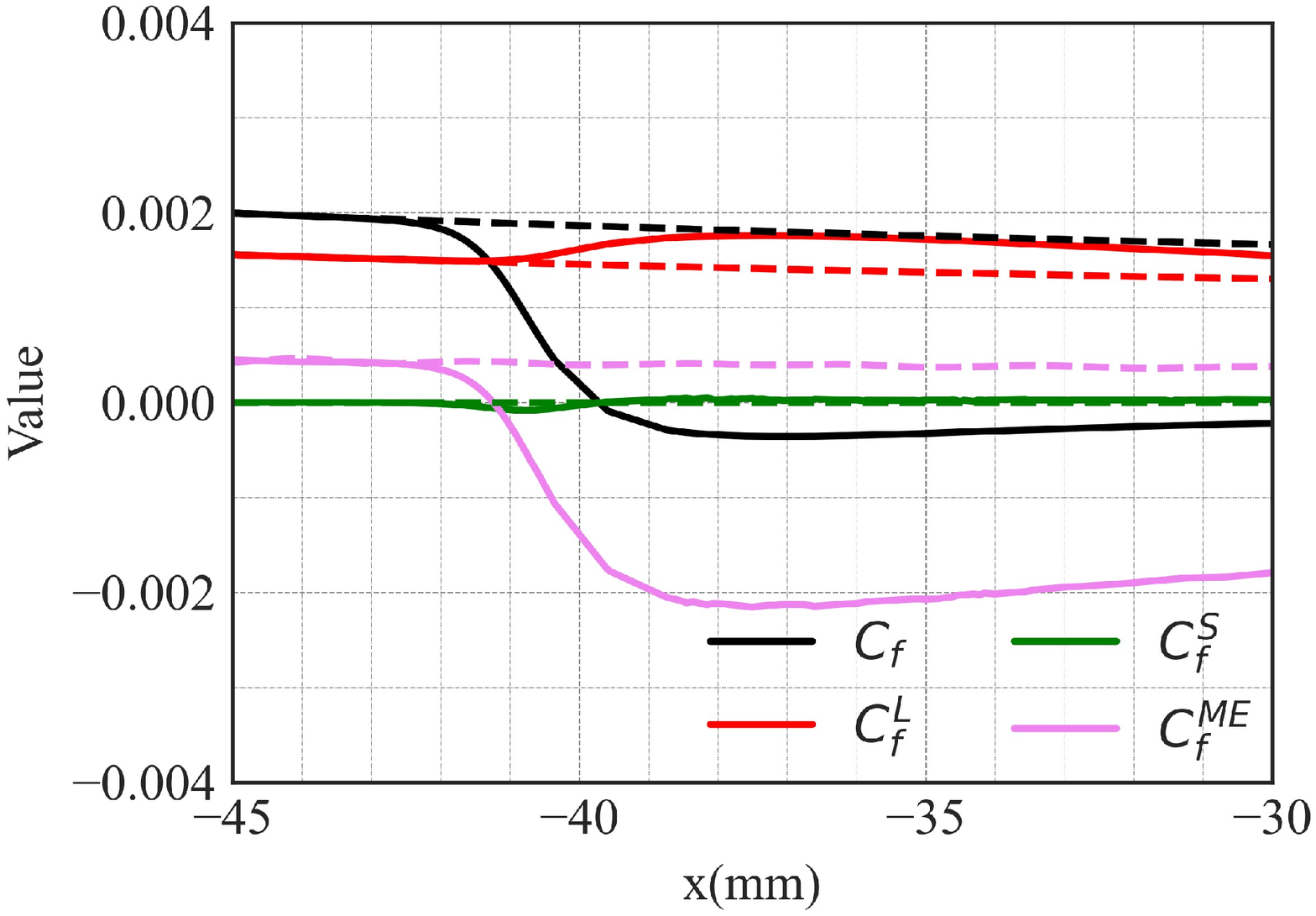}} 
	%	\hspace{-1.5mm}
	\subfigure[\label{subfig:cf2_reatt}]{\includegraphics[width = 0.5\columnwidth]{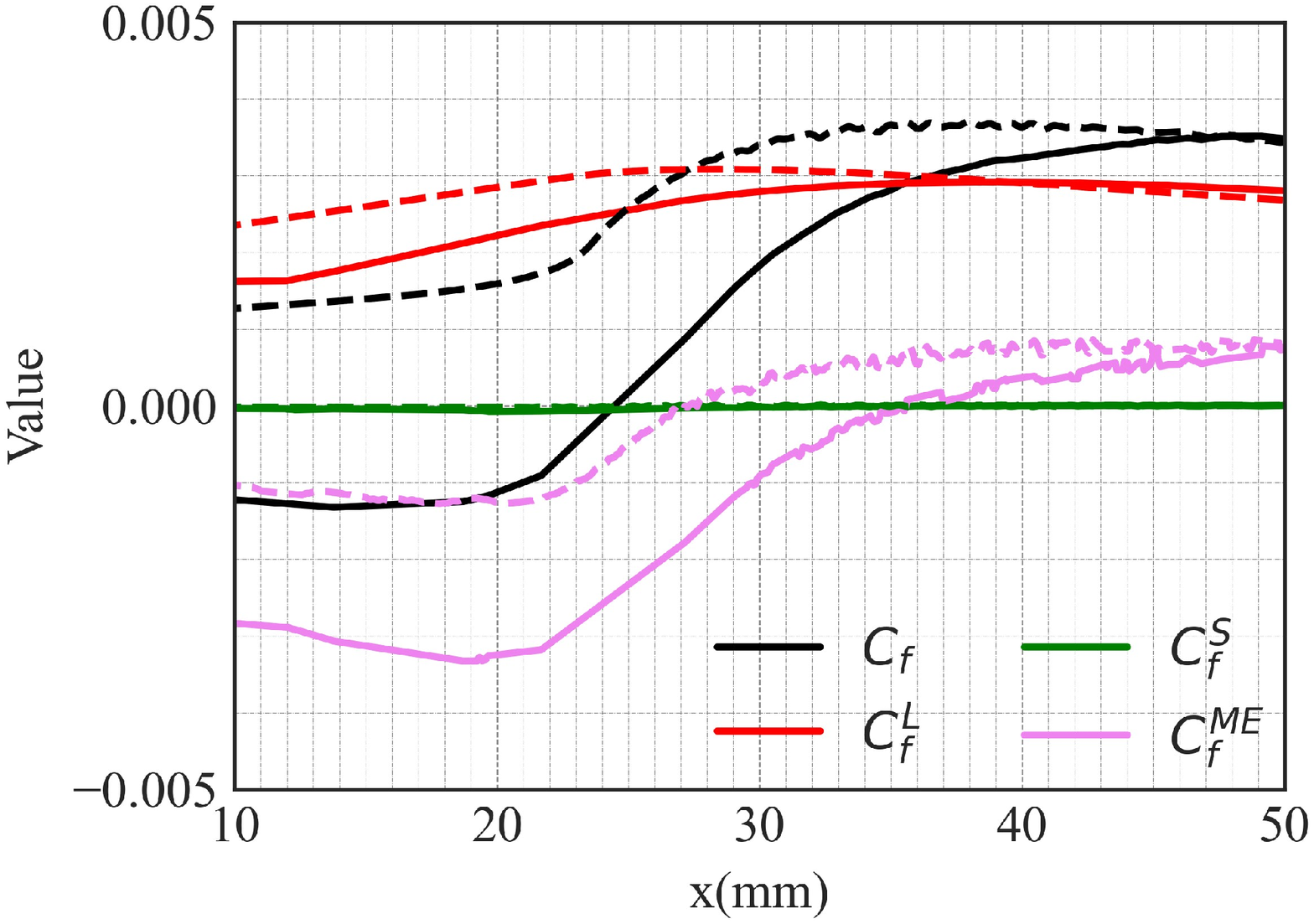}}
	%	\hspace{-1.5mm}		
	\caption{Comparison of friction decomposition of separated/attached states at different streamwise positions (the solid lines are separated states, and dashed lines are attached states.) (a) separation region; (b) reattachment and peak friction region.} \label{fig:cf2_stream}
\end{figure*}

From figure \ref{subfig:cf_sep}, figure \ref{subfig:cf_reatt} and figure \ref{fig:cf_tw}, $C_{f}^{MT}$ and $C_{f}^{P}$ are highly correlated at the separation and reattachment regions of separation states, or curved walls of attachment states. The anti-correlation between them indicates that most of the work by APG  is transformed into kinetic energy transport, which is the transformation of mechanical energy. Then we add $C_{f}^{MT}$ and $C_{f}^{P}$ to obtain the mechanical energy transport term $C_{f}^{ME}$, to further investigate the energy transformation process, as shown in figure \ref{fig:cf2_stream}.

Figure \ref{fig:cf2_stream} clearly shows that the minimum values of friction are caused by the mechanical energy input brought by the separation/reattachment shock wave. The main difference with reference \cite{tang2021aerothermodynamic} is that this paper analyzes from the perspective of energy, whose scale is larger than that of force. Besides, after the deflection of the flow, the viscous dissipation increases obviously, and the mechanical energy input starts to decrease after the reattachment shock wave, so the increase of wall friction is needed to maintain the boundary layer. The coupling of $C_{f}^{L}$ and $C_{f}^{ME}$ leads to the appearance of peak friction. Moreover, the location of peak friction of separation state lies downstream of the attachment state, which results from stronger $C_{f}^{ME}$ in the separation state.

Let's further focus on the energy transformation near the separation point. The friction reduction before the separation point is mainly caused by $C_{f}^{ME}$. At this time, $C_{f}^{L}$ does not change significantly. So the work by the viscous stresses is difficult to dissipate the further input mechanical energy, which leads to the continuous decreased demand for the work done by the wall friction until the negative work, corresponding to the appearance of separation. Therefore, the process that the work by viscous stresses dissipates the input mechanical energy is a key to maintain the boundary layer attachment.

\begin{figure*} 
	%	\hspace{-1.5mm}
%	\centerline
	\subfigure[\label{subfig:reatt_sep}]{\includegraphics[width = 0.5\columnwidth]{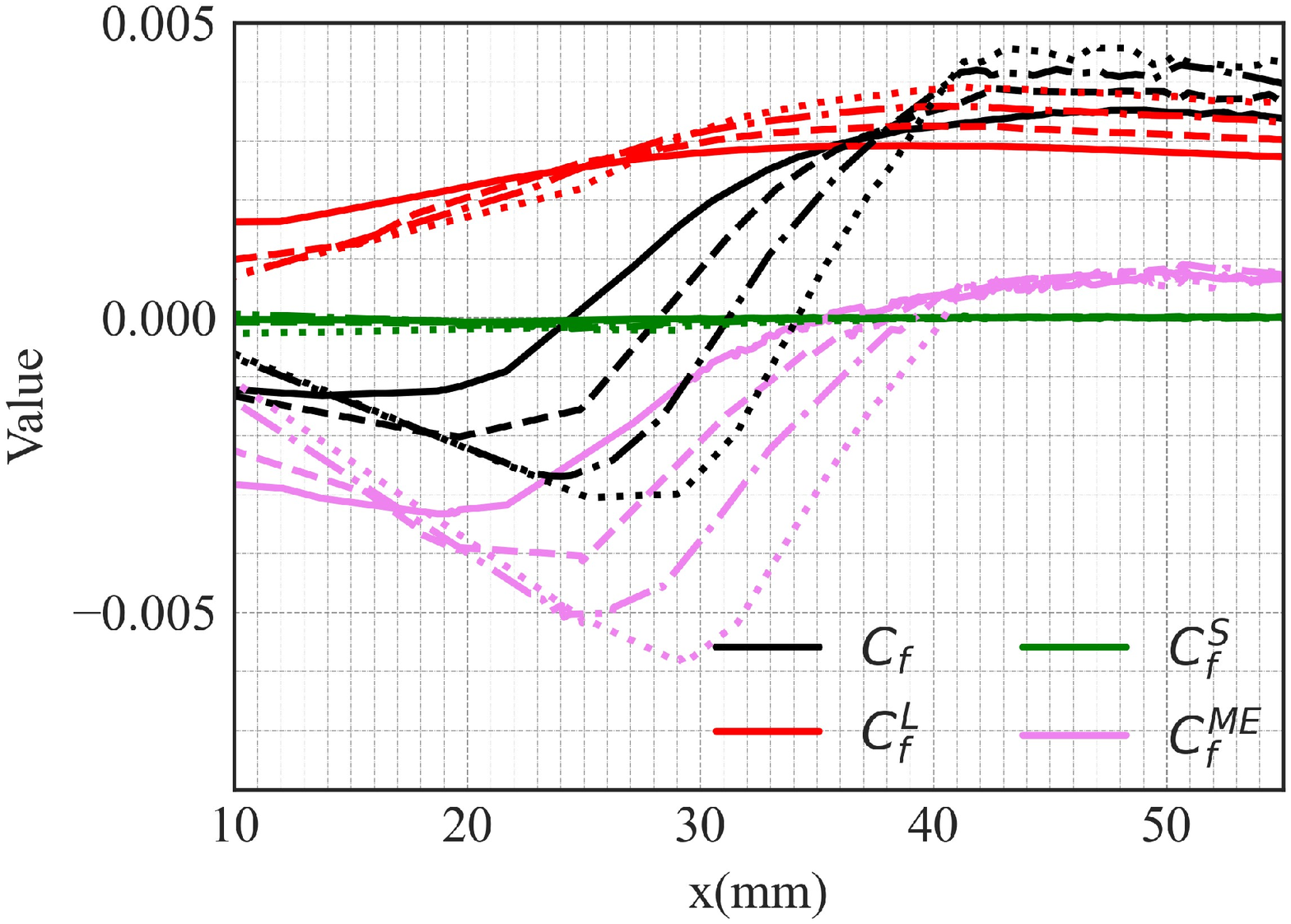}} 
	%	\hspace{-1.5mm}
	\subfigure[\label{subfig:reatt_reatt}]{\includegraphics[width = 0.5\columnwidth]{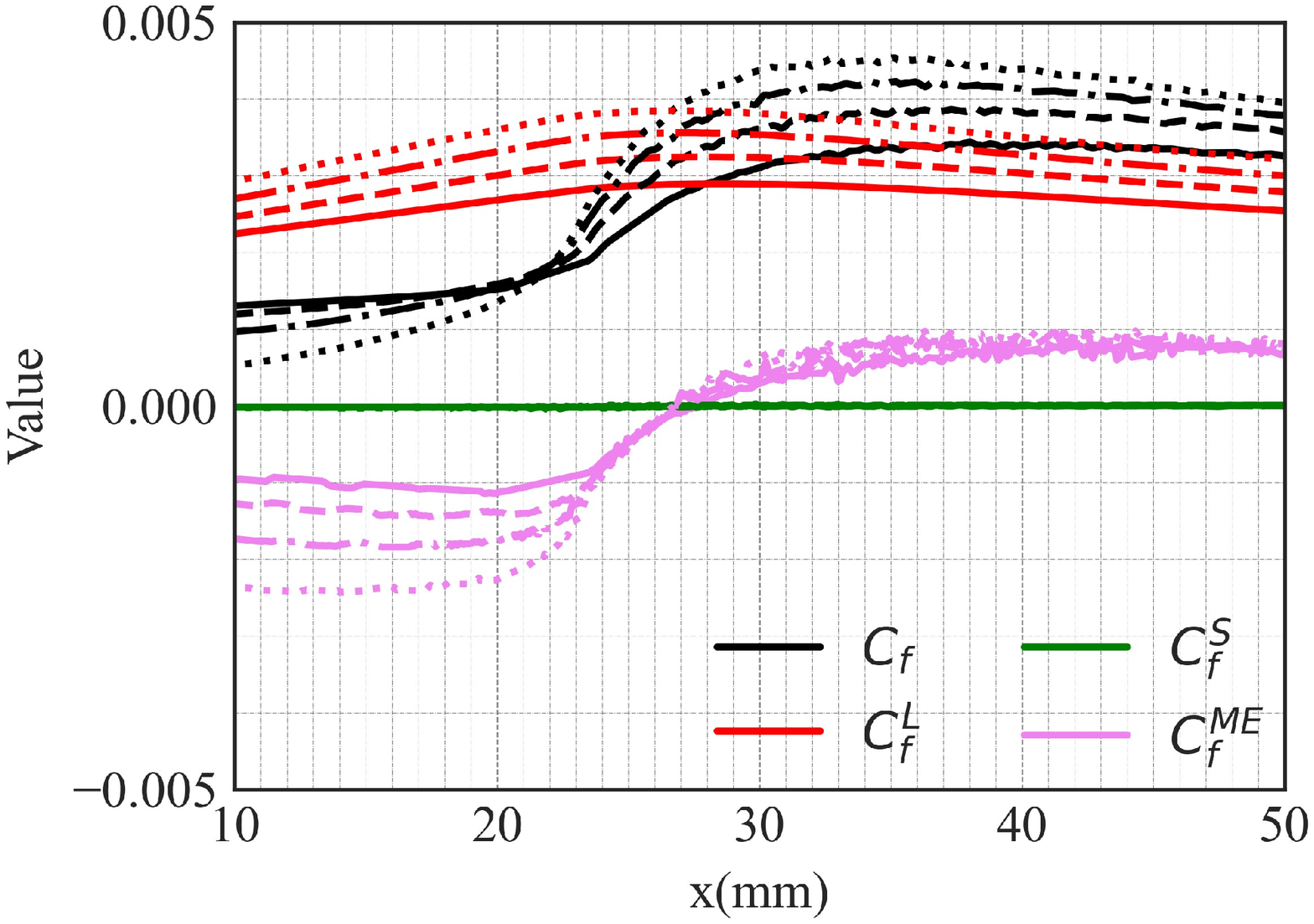}}
	%	\hspace{-1.5mm}		
	\caption{Comparison of enthalpy transport form decomposition in peak friction region (different lines correspond to different corner angles) (a) separation states; (b) attachment states.} \label{fig:reatt_stream}
\end{figure*}

\section {Streamwise evolution of $C_{h}$}

\begin{figure*} 
	%	\hspace{-1.5mm}
%	\centerline
	\subfigure[\label{subfig:ch_bal}]{\includegraphics[width = 0.49\columnwidth]{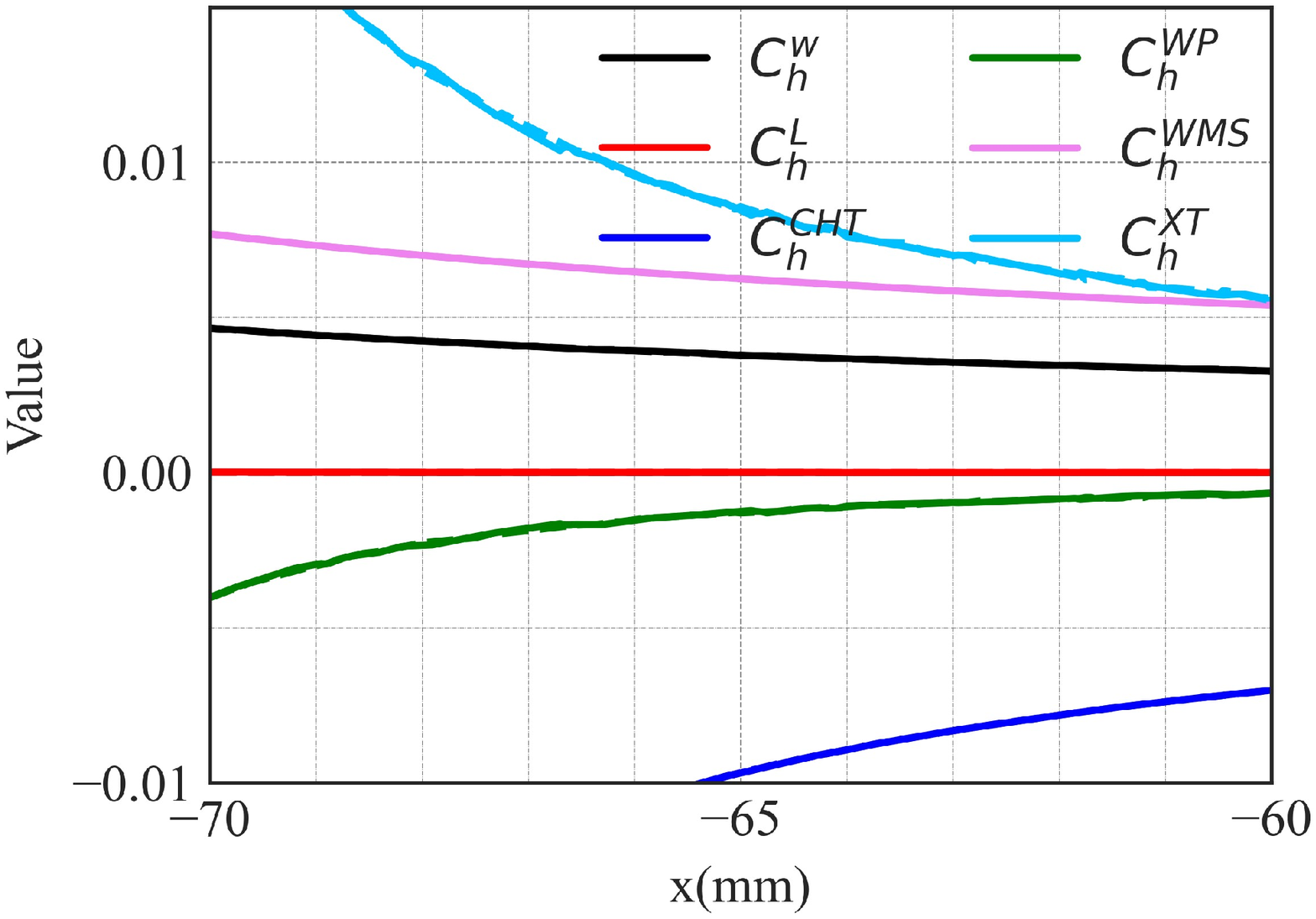}}
	%	\hspace{-1.5mm}		
	\subfigure[\label{subfig:ch_sep}]{\includegraphics[width = 0.5\columnwidth]{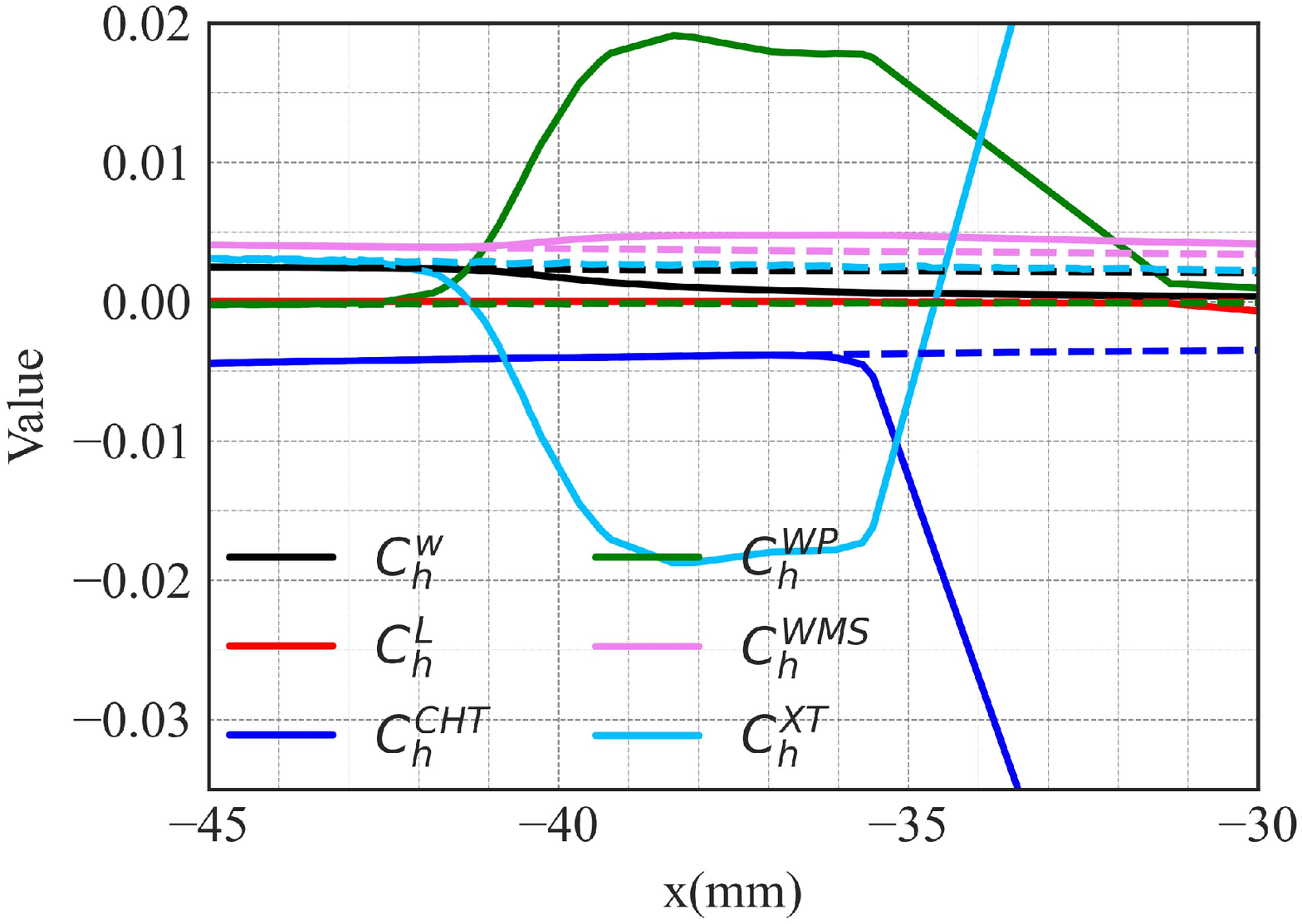}} 
	%	\hspace{-1.5mm}
	\subfigure[\label{subfig:ch_reatt}]{\includegraphics[width = 0.5\columnwidth]{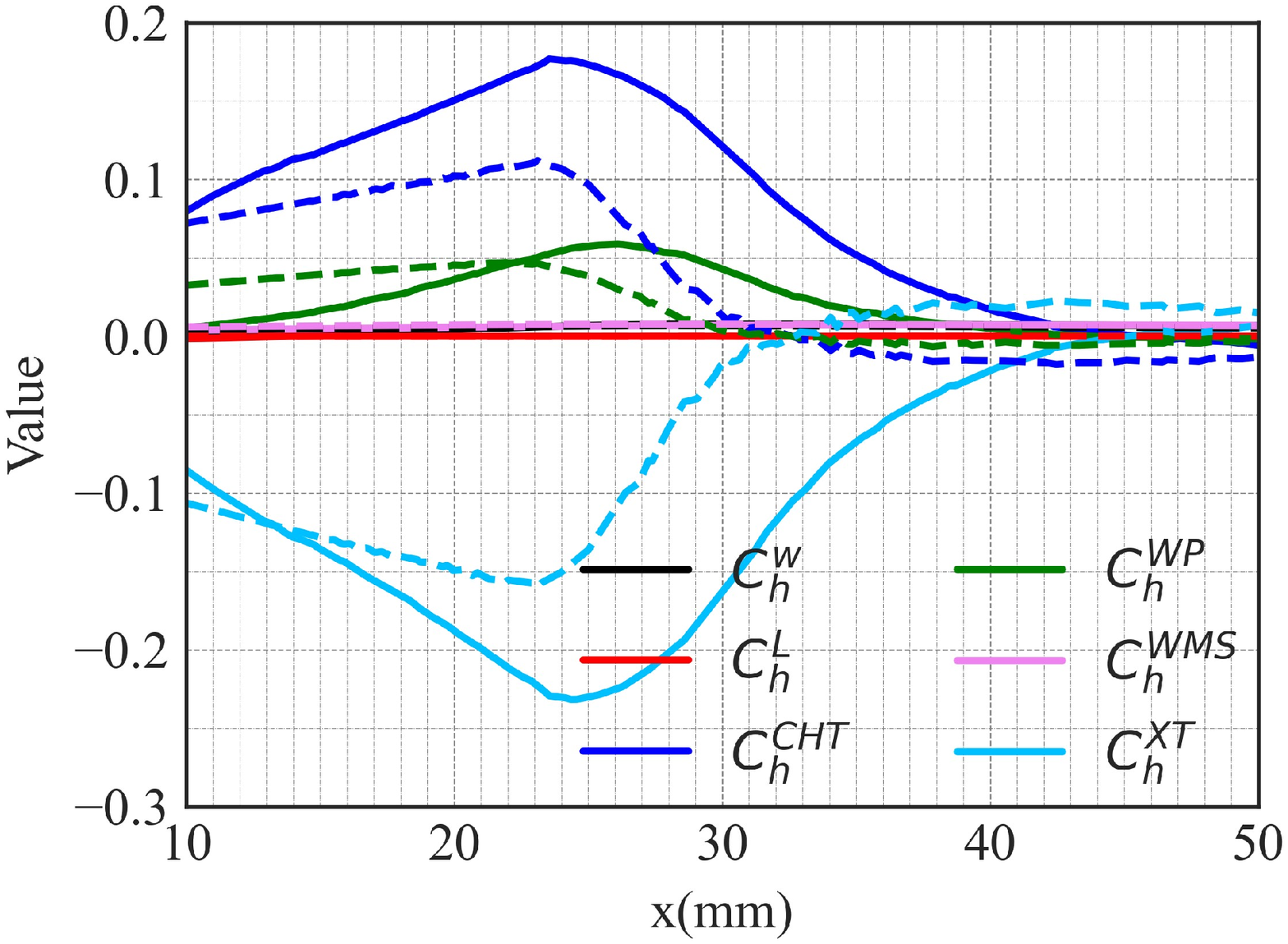}}
	%	\hspace{-1.5mm}		
	\subfigure[\label{subfig:ch_recv}]{\includegraphics[width = 0.5\columnwidth]{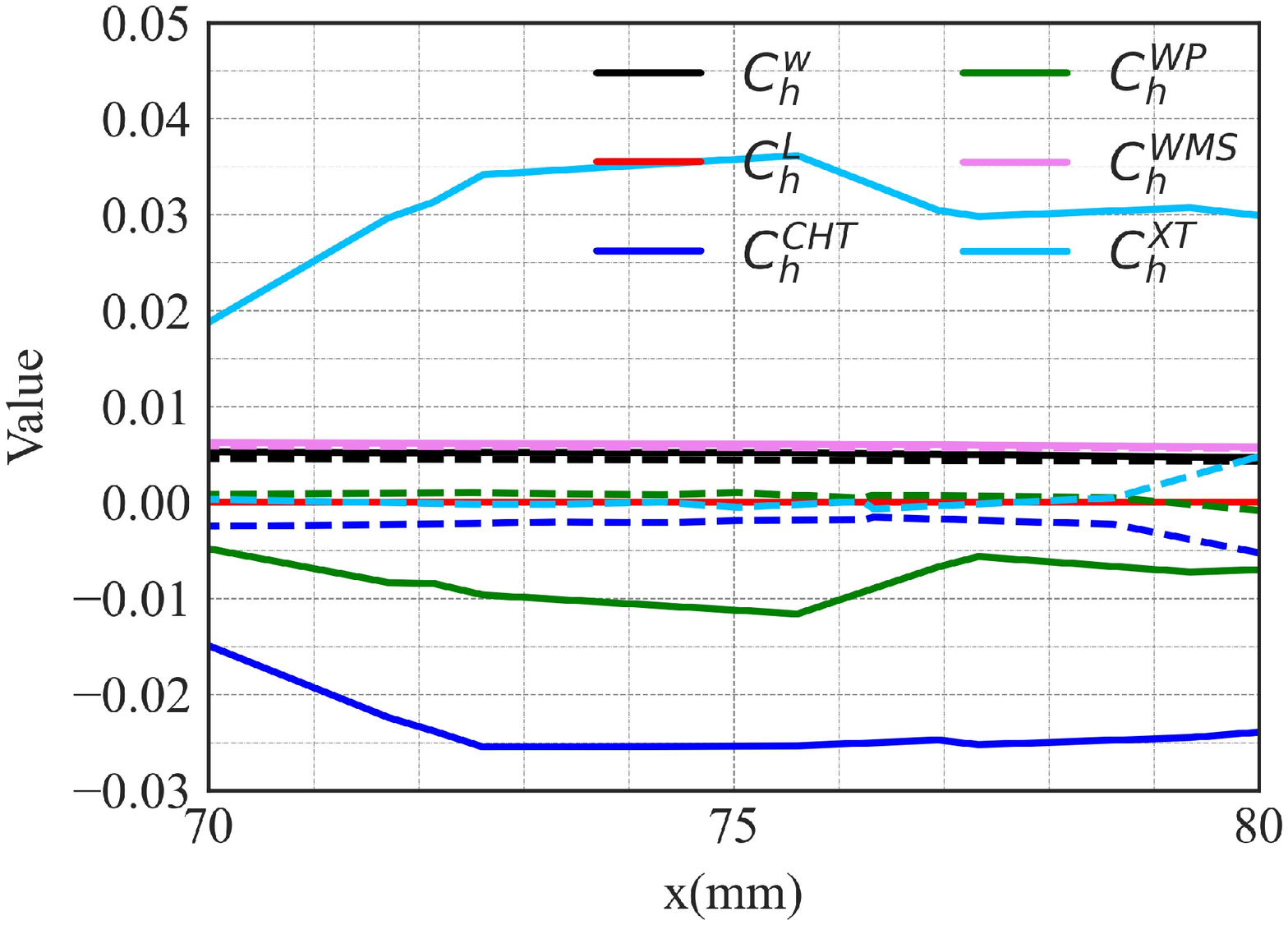}} 
	%	\hspace{-1.5mm}	
	\caption{Comparison of wall heat flux decomposition of separated/attached states at different streamwise positions (the solid lines are separated states, and dashed lines are attached states.) (a) equilibrium flow region; (b) separation region; (c) reattachment and peak friction region; (d) recovery region.} \label{fig:ch_stream}
\end{figure*}

The analysis of wall heat flux decomposition stands in the absolute coordinate system. Similar to the friction decomposition analysis, firstly, the streamwise evolution of each item are analyzed, as shown in figure \ref{fig:ch_stream}. For different regions, the heat transfer $C_{h}^{L}$ is very small and can be neglected. In the equilibrium flow region (figure \ref{subfig:ch_bal}), with the development of the boundary layer, convective heat transport $C_{h}^{CHT}$ , viscous dissipative work $C_{h}^{WMS}$ and enthalpy transport $C_{h}^{XT}$ gradually decrease. Although $C_{h}^{CHT}$ convects the enthalpy outward from the wall to the free flow, the streamwise transport of enthalpy will supplement these parts. It should be noted that $C_{h}^{WMS}$ is higher than $C_{h}$, so it can be inferred that the enthalpy convection along the normal direction is higher than the streamwise accumulation. In other word, $C_{h}^{WMS}$ is not enough to fill the enthalpy deficit caused by $C_{h}^{CHT}$ in the boundary layer. 

For separation area (figure \ref{subfig:ch_sep}), similar to the previous analysis, obvious work by APG caused by the separation shock wave can be found near the separation point. A very interesting phenomenon in this area is that $C_{h}^{CHT}$ and $C_{h}^{XT}$ together lead to energy reduction, rather than anti-correlated between the them. In the reattachment point and peak heat flow region (figure \ref{subfig:ch_reatt}), $C_{h}^{CHT}$ plays a positive role in the increase of wall heat flux due to the impinging of the shear layer in the reattachment process of separation state, and its value is significantly higher than that of the wall heat flux. However, most of this energy is transported downstream. For the attachment state, $C_{h}^{CHT}$ also directs energy to the wall due to the deflection of the flow, but the amount is significantly lower than that of the separation state.

\begin{figure*} 
	%	\hspace{-1.5mm}
%	\centerline
	\subfigure[\label{subfig:ch2_sep}]{\includegraphics[width = 0.5\columnwidth]{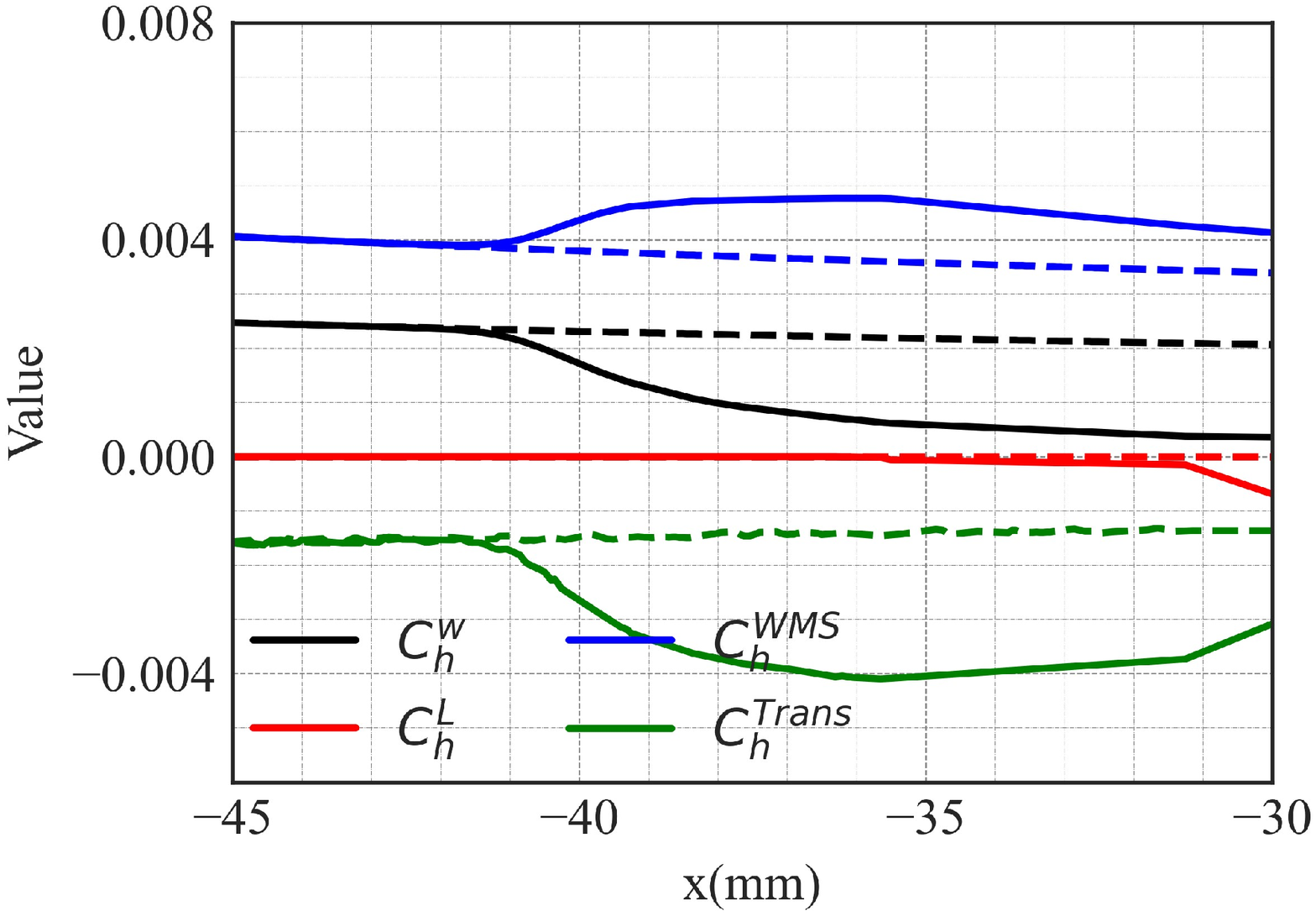}} 
	%	\hspace{-1.5mm}
	\subfigure[\label{subfig:ch2_reatt}]{\includegraphics[width = 0.5\columnwidth]{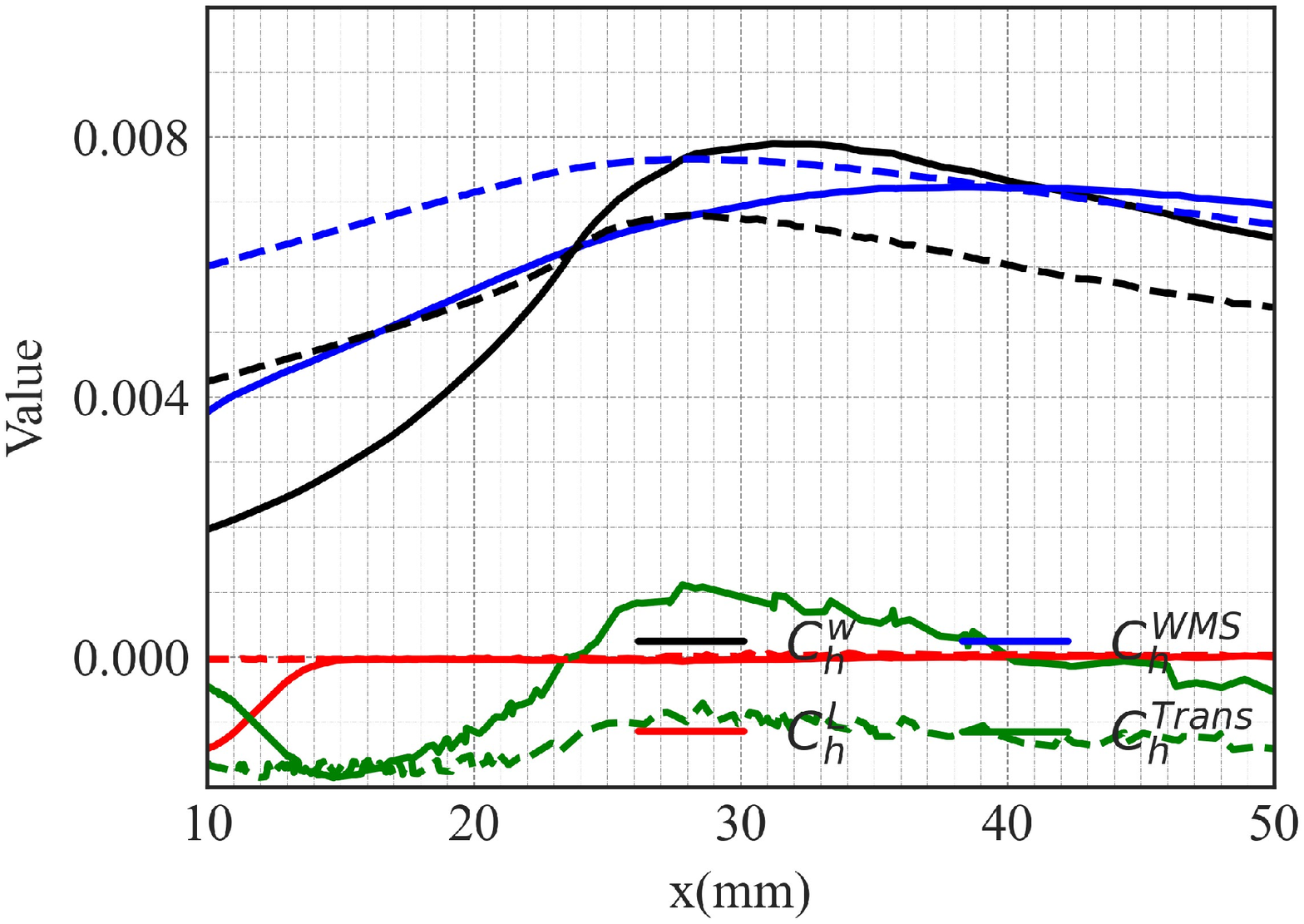}}
	%	\hspace{-1.5mm}		
	\caption{Comparison of wall heat flux decomposition of separated/attached states at different streamwise positions (the solid lines are separated states, and dashed lines are attached states.) (a) separation region; (b) reattachment and peak friction region.} \label{fig:ch2_stream}
\end{figure*}

Since $C_{h}^{CHT}$ and $C_{h}^{XT}$ always exhibit balance characteristics, the two items and the work by APG $C_{h}^{WP}$ are added into energy transport term ($C_{h}^{Trans}$). And the distributions are shown in figure \ref{fig:ch2_stream}.

In separation region (figure \ref{subfig:ch2_sep}), $C_{h}^{WMS}$ even increases to offer more enthalpy, which is transported by $C_{h}^{Trans}$, resulting in the decrease of $C_{h}$. In the region of the peak heat flux (figure \ref{subfig:ch2_reatt}), $C_{h}^{WMS}$ dominant the wall heat flux generation. This result indicates that although the enthalpy brought by the flow impinging is large, little are left locally. For attachment state, the streamwise position of peak wall heat flux and peak $C_{h}^{WMS}$ are nearly the same. However, the position of peak heat flux of the separation state is affected by $C_{h}^{Trans}$, which  is also a key parameter for the classical peak heat flow prediction model\cite{simeonides1994experimental,simeonides1995experimental}.

\begin{figure*} 
	%	\hspace{-1.5mm}
%	\centerline
	\subfigure[\label{subfig:phi_sep}]{\includegraphics[width = 0.5\columnwidth]{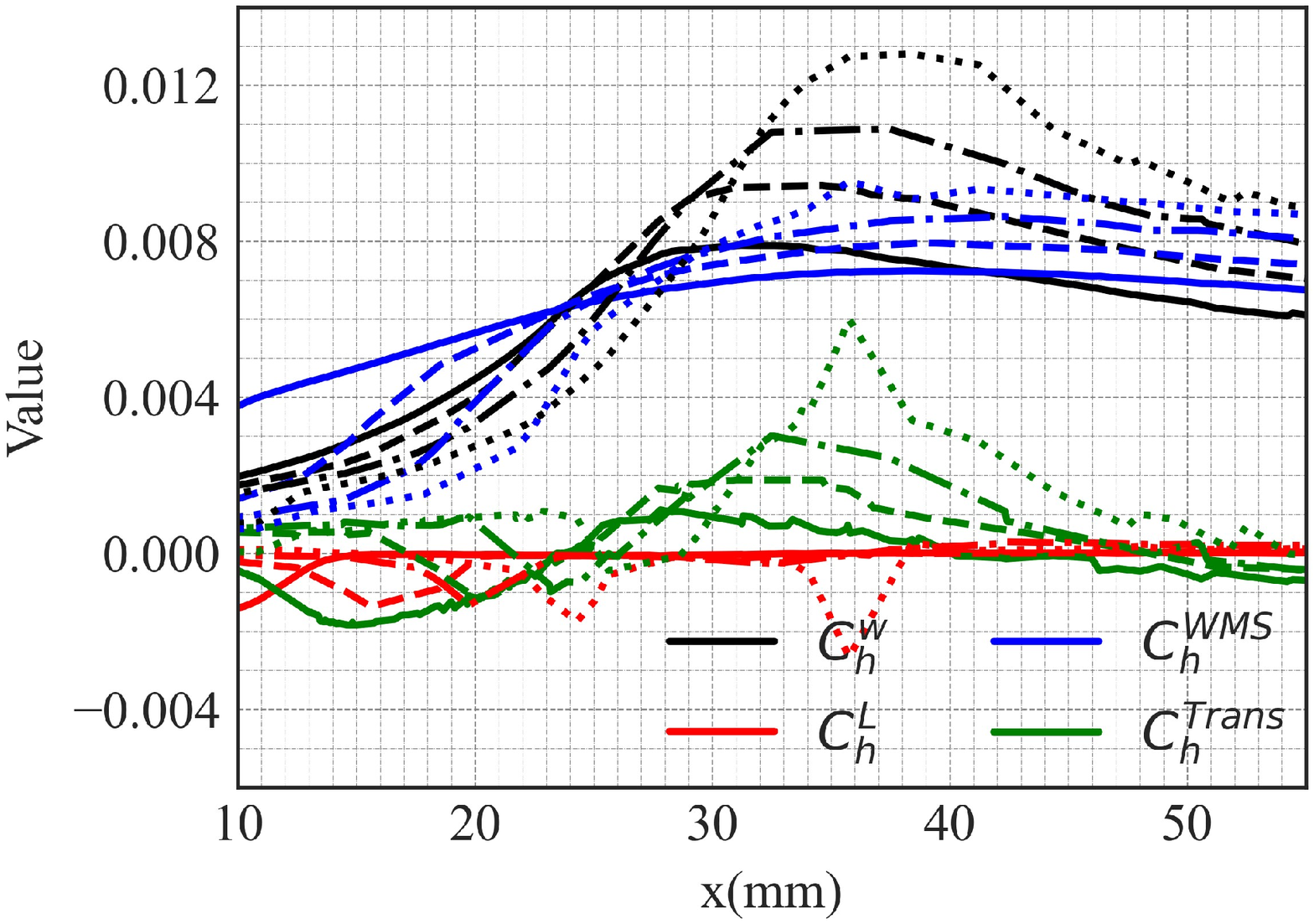}} 
	%	\hspace{-1.5mm}
	\subfigure[\label{subfig:phi_att}]{\includegraphics[width = 0.5\columnwidth]{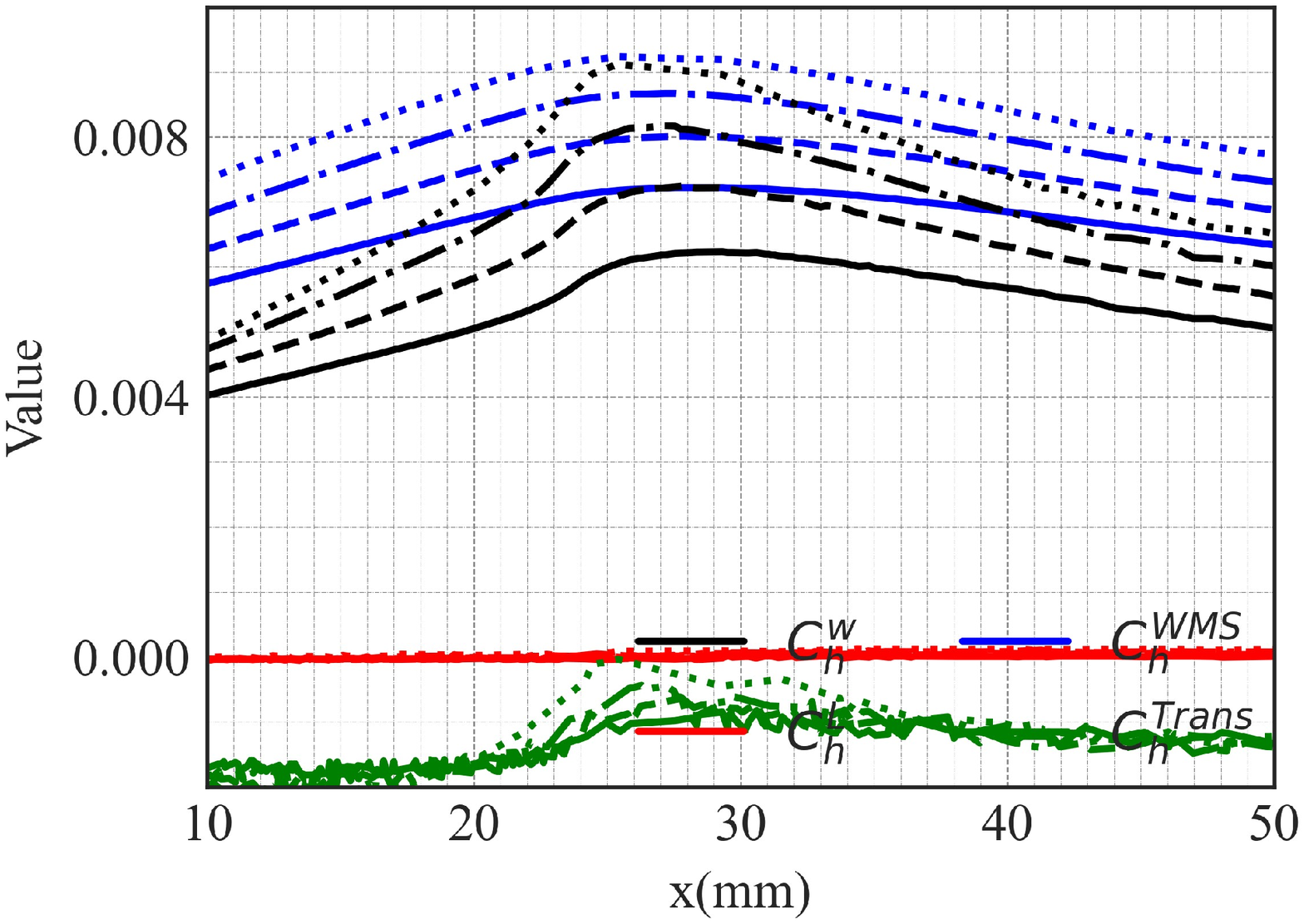}}
	%	\hspace{-1.5mm}		
	\caption{Comparison of decomposition of energy transport forms in peak heat flow region (different lines correspond to different corner angles) (a) separation states; (b) attachment states.} \label{fig:phi_stream}
\end{figure*}

\begin{figure*} 
	%	\hspace{-1.5mm}
%	\centerline
	\subfigure[\label{subfig:tw_sep}]{\includegraphics[width = 0.5\columnwidth]{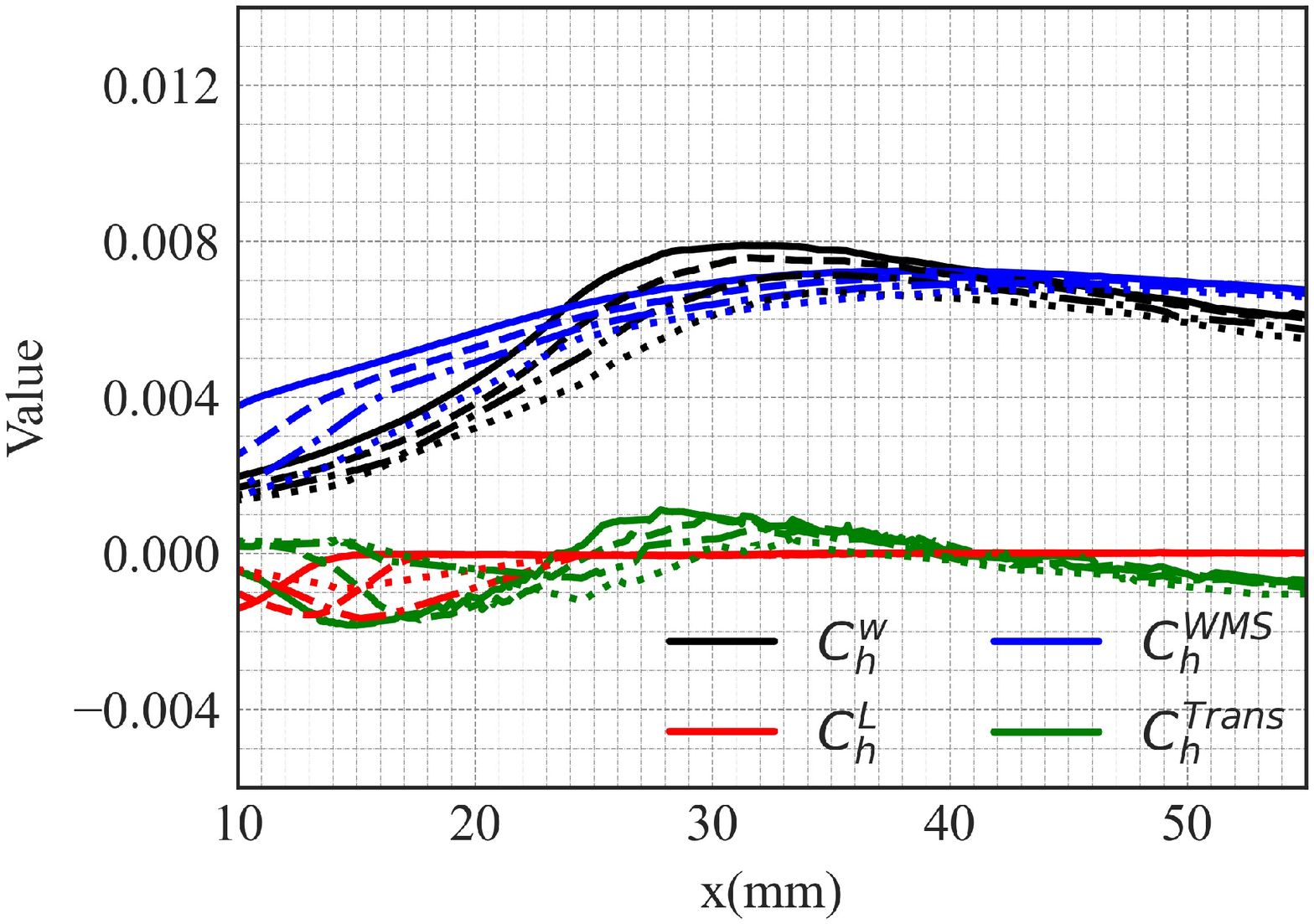}} 
	%	\hspace{-1.5mm}
	\subfigure[\label{subfig:tw_att}]{\includegraphics[width = 0.5\columnwidth]{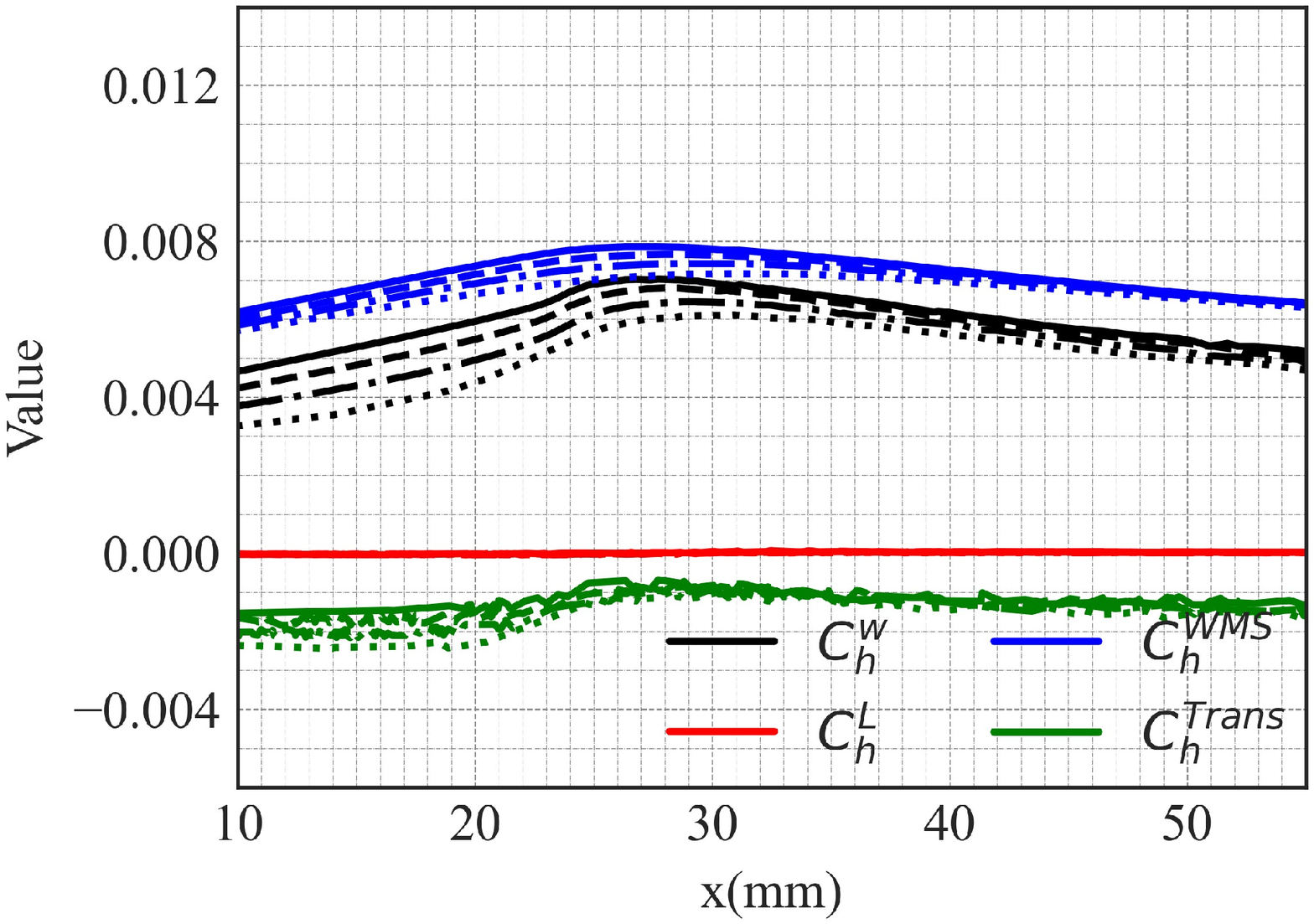}}
	%	\hspace{-1.5mm}		
	\caption{Comparison of wall heat flux decomposition of separated/attached states at different streamwise positions (the solid lines are separated states, and dashed lines are attached states.) (a) separation states; (b) attachment states.} \label{fig:tw_stream}
\end{figure*}

\section {The Reynolds analogy factor}

Before discussing Reynolds's comparison relation, using the boundary layer thickness $\delta$ as the upper bound of the integral of equation \ref{eq:step1_ch}, i.e., 1-fold integration is carried out, then we can obtain

\begin{align}\label{eq:ch_dec1}
	C_{h}=&\frac{q_{y, \delta}}{\rho_{\infty} u_{\infty} c_{p}\left(T_{a w}-T_{w}\right)}-\frac{\left.\overline{\rho h v}\right|_{\delta}}{\rho_{\infty} u_{\infty} c_{p}\left(T_{a w}-T_{w}\right)} \\\nonumber
	&+\frac{1}{\rho_{\infty} u_{\infty} c_{p}\left(T_{a w}-T_{w}\right)} \int_{0}^{\delta} \overline{\tau_{\imath, j} \frac{\partial u_{\imath}}{\partial x_{j}}} d y \\\nonumber
	&+\frac{1}{\rho_{\infty} u_{\infty} c_{p}\left(T_{a w}-T_{w}\right)} \int_{0}^{\delta}\left(u_{j} \frac{\partial p}{\partial x_{j}}-\frac{\partial \overline{\rho h u}}{\partial x}\right) d y
\end{align}

The shortage of equation \ref{eq:ch_dec1} is that the contribution from Reynolds shear stresses to wall heat flux generation is eliminated. However, 
the viscous stress work term $C_{h}^{WMS,1}$ in equation \ref{eq:ch_dec1} is very close to the viscous dissipation term in equation \ref{eq:step2}. $C_{h}^{WMS,1}$ is further expanded and can be expressed as 

\begin{align}\label{eq:ch_wms1}
	\int_{0}^{\delta} \tau_{i, j} \frac{\partial u_{i}}{\partial x_{j}} d y=\underbrace{\int_{0}^{\delta}\left(\overline{\tau_{x x} \frac{\partial u}{\partial x}}+\overline{\left.\tau_{y y} \frac{\partial v}{\partial y}\right)} d y\right.}_{C_{h}^{M S \_ Dilat}}+\underbrace{\int_{0}^{\delta} \overline{\tau_{x y}\left(\frac{\partial u}{\partial y}+\frac{\partial v}{\partial x}\right)} d y}_{C_{h}^{M S \_ Dissip }}
\end{align}

The difference between ${C_{h}^{M S \_ Dissip }}$ and ${C_{f}^{L}}$ is the term $\overline{\tau_{x y} \frac{\partial v}{\partial x}}$. In order to investigate the effect of different terms on Reynolds analogy factor, the 1-fold and 2-fold integration of heat flux decomposition are both analyzed. As the wall heat flux is close to 0 in the separation bubble, the attachment states are analyzed below to obtain more information.

\begin{figure*} 
	%	\hspace{-1.5mm}
%	\centerline
	\subfigure[\label{subfig:ra1}]{\includegraphics[width = 0.5\columnwidth]{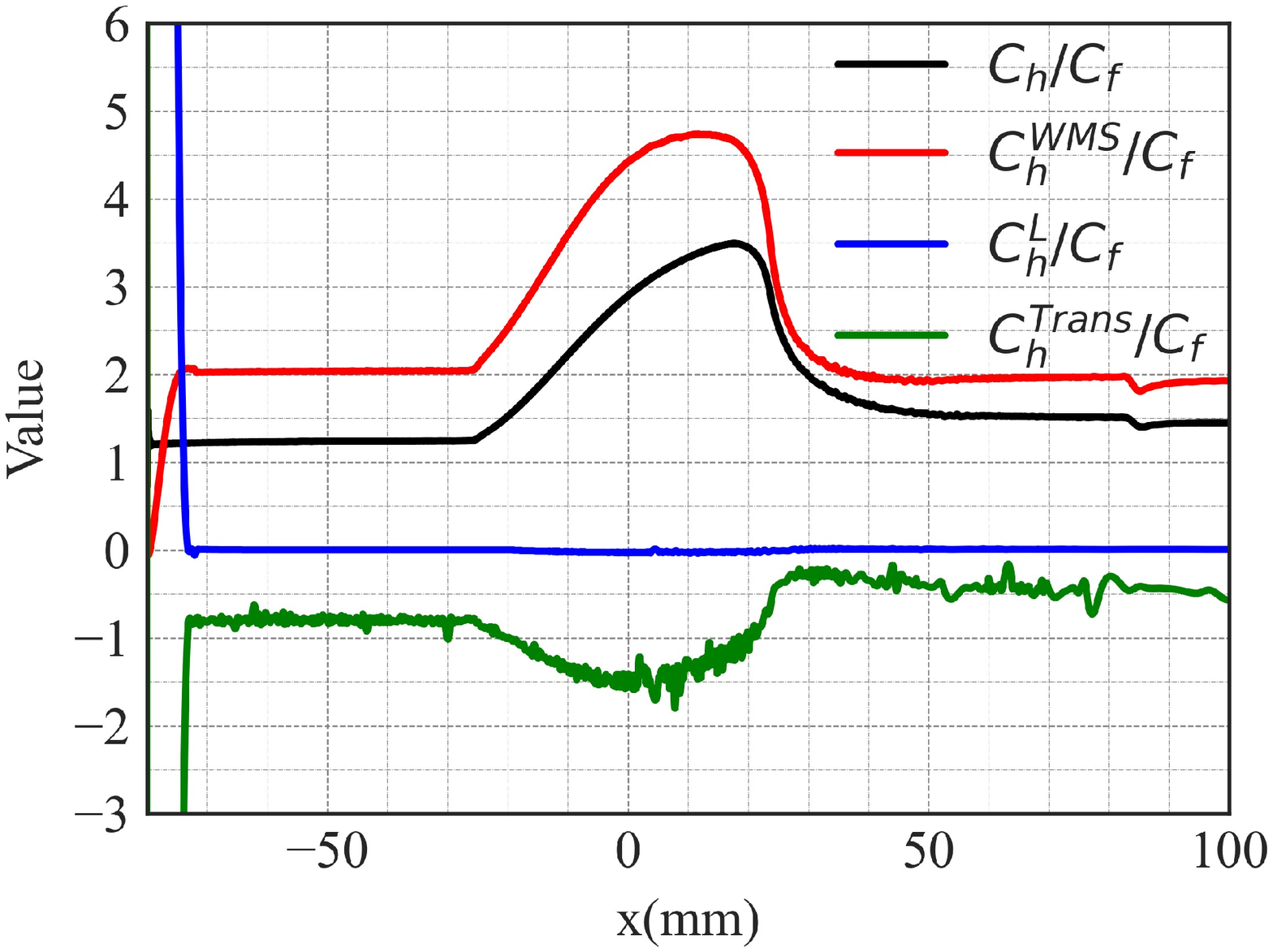}} 
	%	\hspace{-1.5mm}
	\subfigure[\label{subfig:ra2}]{\includegraphics[width = 0.5\columnwidth]{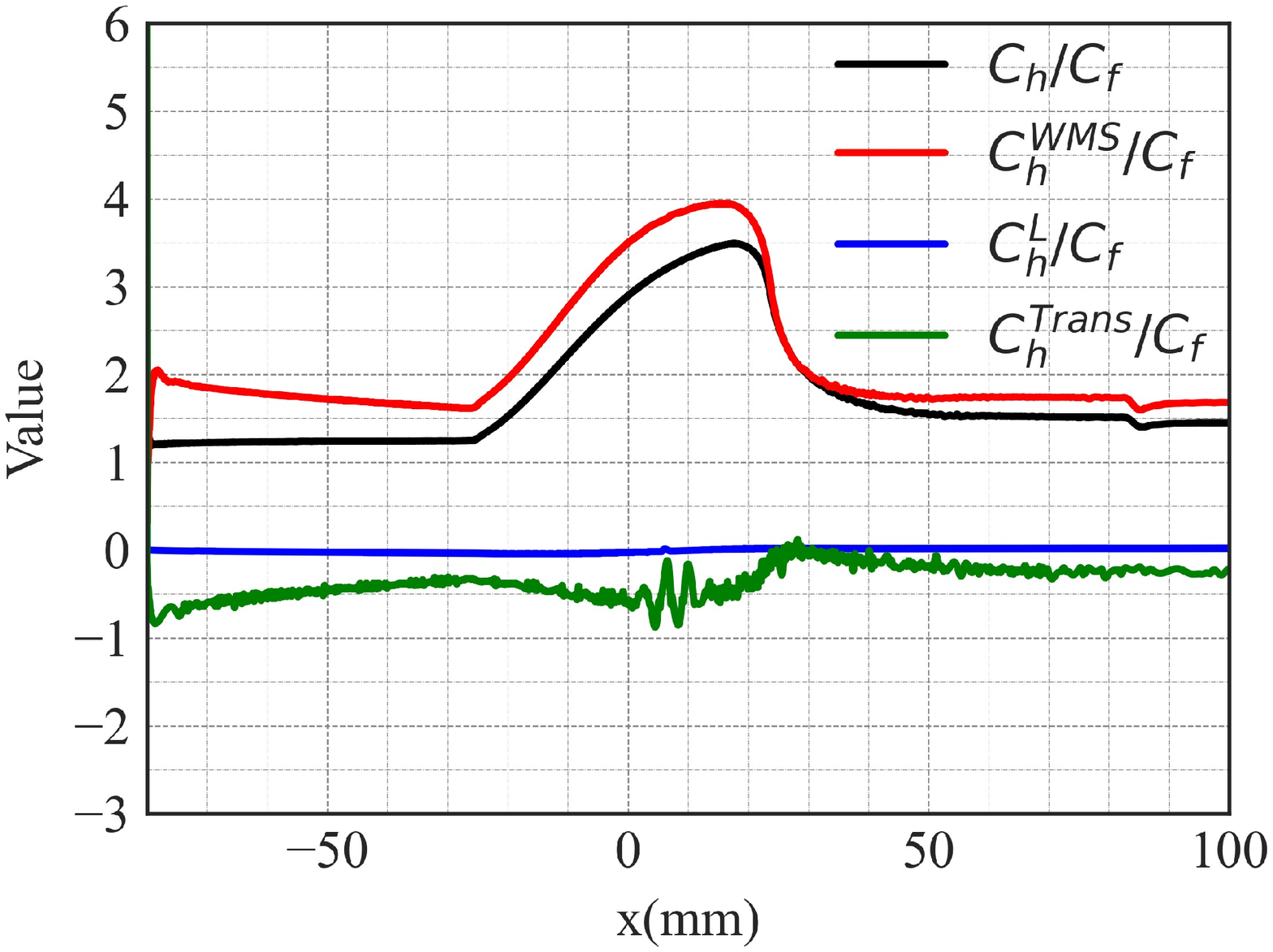}}
%	\hspace{5.0mm}		
	\centerline{\subfigure[\label{subfig:apg}]{\includegraphics[width = 0.5\columnwidth]{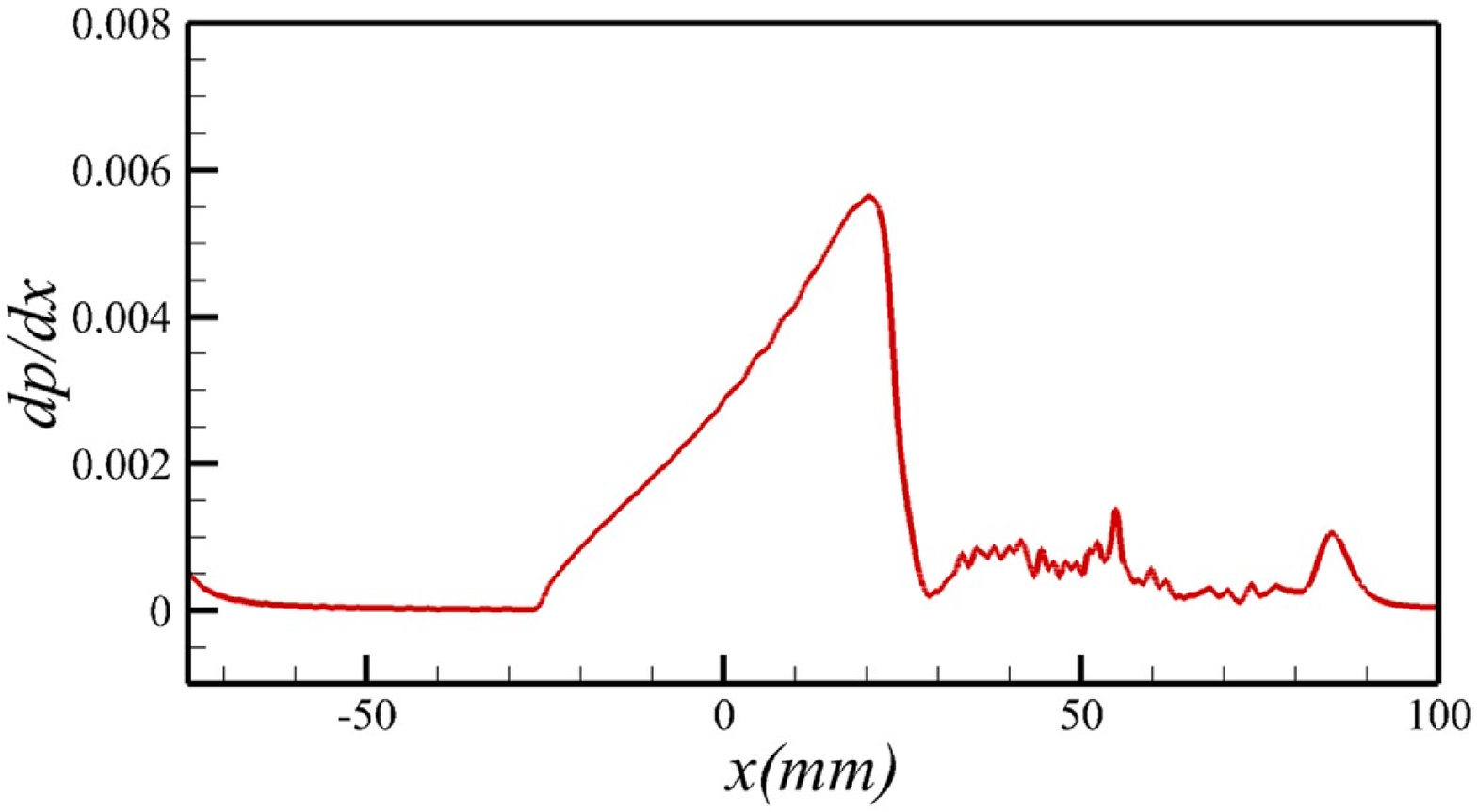}}}
	
	\caption{Streamwise distributions of various Reynolds analogy factors of friction and heat flux decomposition. (a) 1-fold integration; (b) 2-fold integration; (c) streamwise distribution of APG.} \label{fig:ra}
\end{figure*}

As shown in figure \ref{fig:ra}, the Reynolds analogy factor $s= 2 {C_{h}} / {C_{f}} $ is constant when the boundary layer is not affected by APG ($x < 25mm$). While $s$ returns to constant at $x > 50mm$, which locates downstream of the end of the curved wall ($x \approx 25mm$), where APG approaching 0. And there $s$ rises by about 20\% compared with the flow on the flat plate. On the curved wall, $s$ increases with the increase of APG (figure \ref{subfig:apg}), which is consistent with Wenzel's study. As shown from figure \ref{fig:dec_vf}, the increase comes from the simultaneous decrease of friction and increase of heat flux. From previous analysis, the dominant iterms include $C_{f}^{L}$ and $C_{f}^{P}$. And it is $C_{f}^{P}$ that dominates the opposite tendency of $C_{f}$ and $C_{h}$ with the effect of APG. 

From further comparison of the terms of the heat flux decomposition, $C_{h}^{L}$ is quite small and can be ignored in both 1-fold (figure \ref{subfig:ra1}) and 2-fold integrations (figure \ref{subfig:ra2}). The difference of $s$ between upstream and downstream of the curved wall mainly results from the change of energy transport, and the energy transport weakens at the downstream. In addition, $C_h^{WMS} / {C_f}$ of 2-fold integration decreases along streamwise direction on the flat plate, compared with the result of 1-fold integration. Such tendency reveals that the proportion of viscous dissipation and energy transport near the wall decreases with the development of the boundary layer. However, the results of 1-fold and 2-fold integrations both become nearly constant at $x > 50mm$. Meanwhile, $C_h^{Trans} / {C_f}$ also becomes nearly constant of about 0. The results reveals that streamwise variation of velocity and temperature profiles of boundary layer is small, compared with that on the flat plate.

\begin{figure*} 
	%	\hspace{-1.5mm}
%	\centerline
	\subfigure[\label{subfig:ra_terms1}]{\includegraphics[width = 0.5\columnwidth]{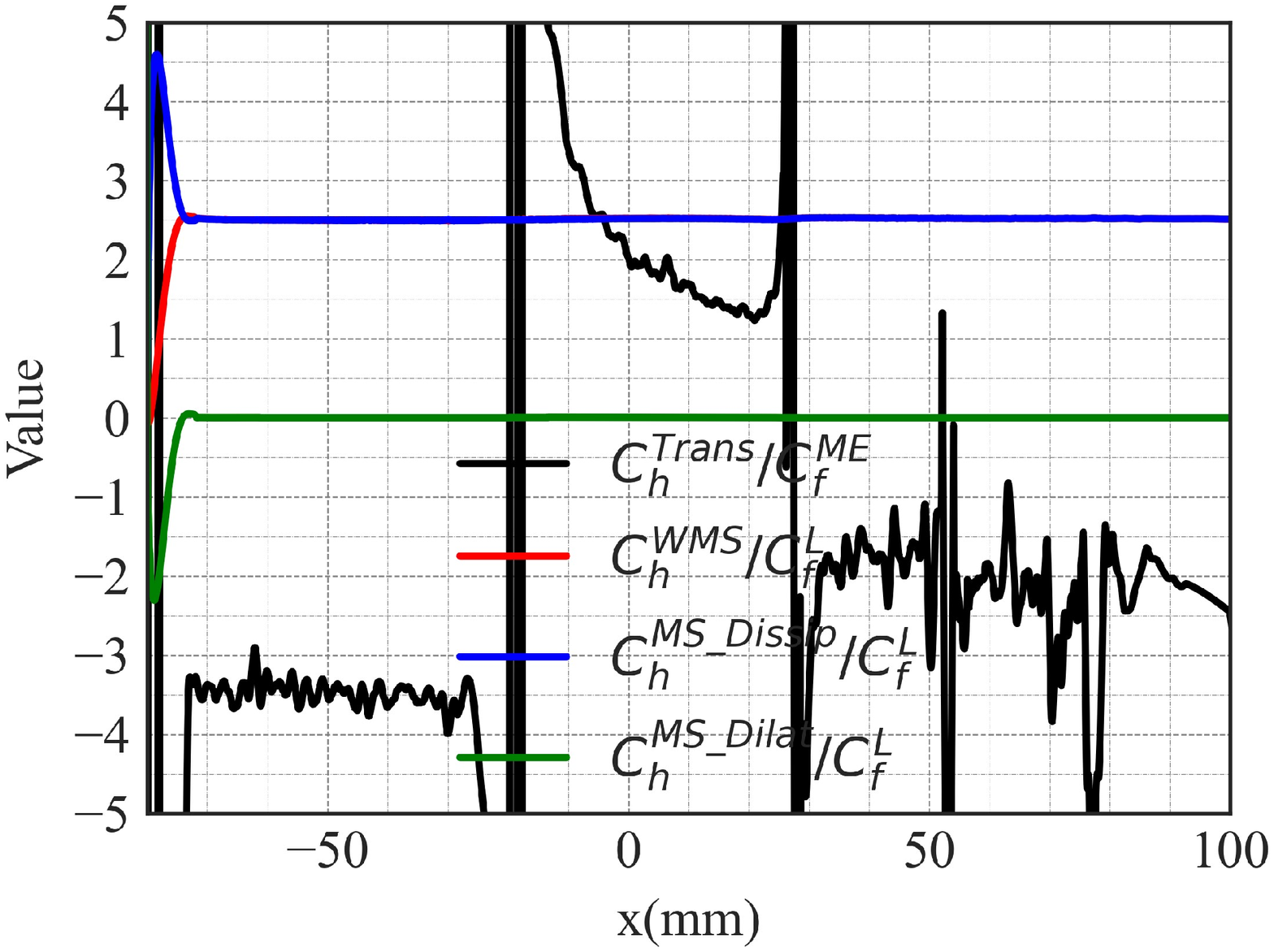}} 
	%	\hspace{-1.5mm}
	\subfigure[\label{subfig:ra_terms2}]{\includegraphics[width = 0.5\columnwidth]{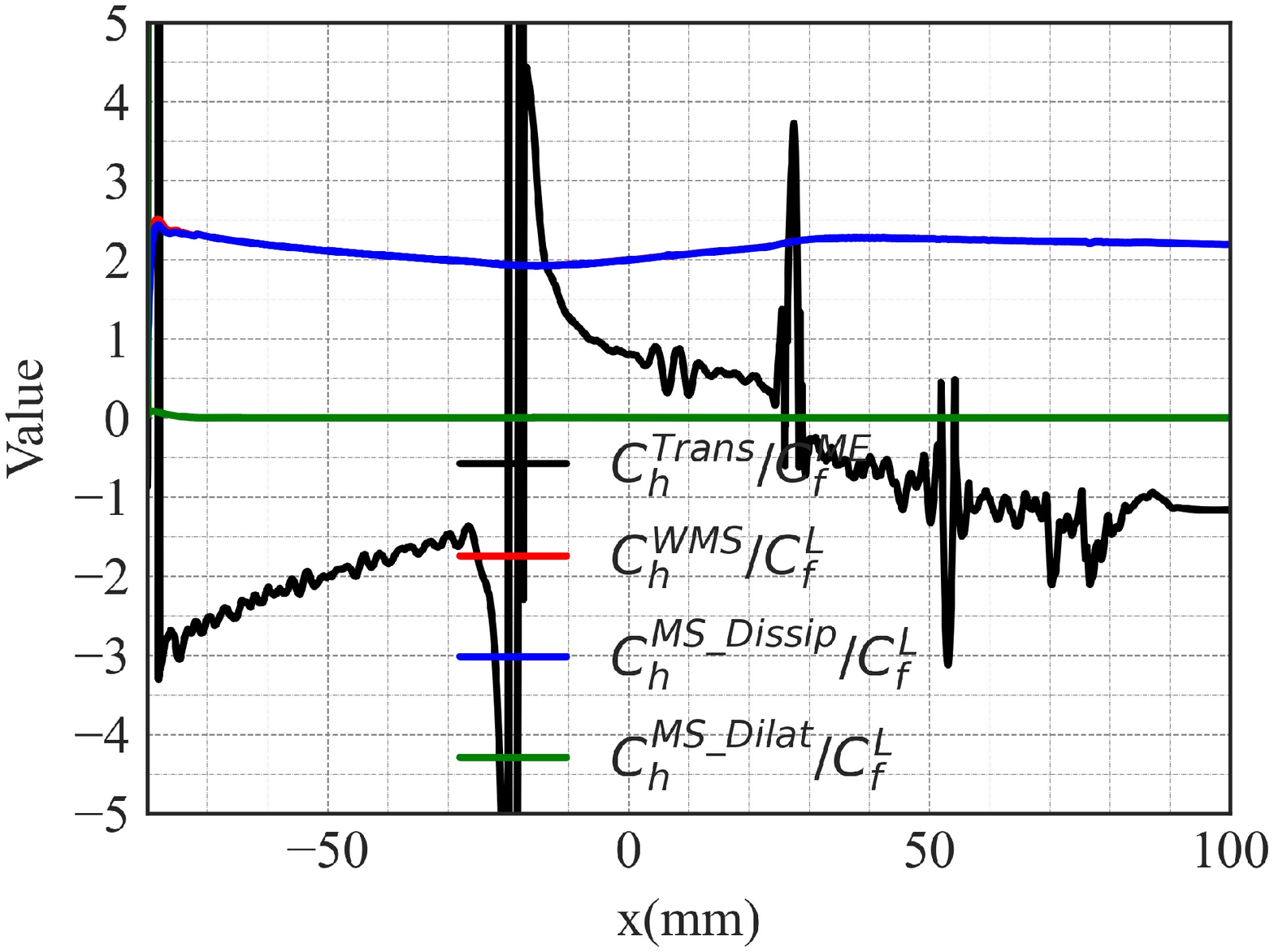}}
	%	\hspace{-1.5mm}		
	\caption{Streamwise distributions of the correlations between the terms of friction and heat flux decomposition. (a) 1-fold integration; (b) 2-fold integration.} \label{fig:ra_terms}
\end{figure*}

Now we investigate the relationship of the terms between heat flux decomposition and friction decomposition. $C_h^{Trans}$ and ${C_f^{ME}}$ exhibits a correlation before and after the curved wall (figure \ref{fig:ra_terms}), representing the inversely proportion of the energy outflow and the involved the mechanical energy during the boundary layer development (the energy deficit in absolute coordinate system and the kinetic energy gain in relative coordinate system of boundary layer). And the difference between before and after the curved wall is induced by temperature rise and velocity decrease, and the relation between them can be expressed as 

\begin{align}\label{eq:ra1}
	\frac{C_{h}^{\text {Trans }}}{C_{f}^{M E}} \overset{\frac{\partial p}{\partial x_{j}} \approx 0}{=} \frac{\cos \varphi}{2 \sqrt{\operatorname{Pr}}}\frac{\int_{0}^{\delta}\left[-\frac{\partial \overline{\rho h u}}{\partial x}-\frac{\partial \overline{\rho h v}}{\partial y}\right] d y}{\left[\int_{0}^{\delta}\left(\tilde{u}-u_{\infty} \cos \varphi\right)\left(\frac{\partial \overline{\rho u u}}{\partial x}+\frac{\partial \overline{\rho u v}}{\partial y}\right)\right] d y} \approx - \frac{\gamma-1}{2} \frac{\cos \varphi}{2 \sqrt{\mathrm{Pr}}} f\left(q_{w}\right) M a_{e}^{2}
\end{align}

The work by visous stresses ${C_{h}^{WMS}}$ consists of ${C_{h}^{M S \_ Dissip }}$ and ${C_{h}^{M S \_ Dilat }}$ (equation \ref{eq:ch_wms1}). From figure \ref{fig:ra_terms}, ${C_{h}^{M S \_ Dilat }}$ has little effect, and ${C_{h}^{WMS}}$ is almost all dissipated by viscous stresses. From the streamwise distribution of the terms, ${C_{h}^{WMS}}$ of the 1-fold integration equals ${C_{f}^{L}}$ on the flat plate, indicating $\partial u / \partial y=\partial v / \partial x$. When the flow is subjected to APG, ${C_{h}^{WMS}} / {C_{f}^{L}}$ decreases slightly, and the lower value of ${C_{h}^{WMS}} / {C_{f}^{L}}$ maintains after the influence of APG, i.e., $\partial u / \partial y<\partial v / \partial x$, and the relation maintains even at the recovery region. Besides, the relation $\partial u / \partial y<\partial v / \partial x$ indicates that the evolution of velocity on the normal direction is slower than that of viscous dissipation, resulting in lower ejectment speed of enthalpy by normalwise speed than the enthalpy generation by viscous dissipation, which is one of the reasons for the increase of Reynolds analogy factor.

\section {Conclusions}

Based on FIK and RD decomposition, a combined decomposition method with clear physical interpretation and suitable for arbitrary surface is proposed. The derivation of this method bases on the physical process that the boundary layer is maintain by energy input from the wall friction, and the wall heat flux results from the normalwise redistribution of temperature by normalwise velocity. The aim is to associate friction with heat flux from the perspective of energy transformation, and explore a fast prediction method of peak heat flux  by Reynolds analogy. Based on this method, the aerothermodynamic characteristics of a set of curved compression ramp flows with bistable states of separation/attachment are analyzed.

Wall friction is the result of viscous dissipation and work by APG. Moreover, viscous dissipation and work by APG are correlated. In the relative coordinate system where the flow is still, APG will inject energy into the boundary layer, a large proportion of which is converted into kinetic energy, while the other part needs to be dissipated by viscous stresses, otherwise the flow will be separated. This process is similar to the free interaction theory, that is, the flow separation happens when the energy input from the local strong APG cannot be dissipated by the local viscous stresses.

The wall heat flow is dominated by viscous dissipation. In the separation-reattachment process caused by shock wave/boundary layer interaction, although the energy brought by the shear layer impinging is high, most of it is transported downstream and has little influence locally. However, the effect of energy transport on peak heat flux generation increases with the increase of ramp angle, i.e., the increase of APG.

The Reynolds comparison factor will increase with APG. The reason is that the work by APG leads to the decrease of friction, and increase of viscous dissipation resulting in the increase of heat flux. Besides, the combined decomposition method gives the possibility of quantifying the effect of APG on Reynold analogy factor from the perspective of energy.

Since the generation of peak friction and heat flux is dominated by viscous dissipation, reducing viscous dissipation is a potential way to achieve the control of friction and heat flux simultaneously. However, reducing viscous dissipation will lead to a decrease in the ability of the boundary layer to resist APG, which may be one reason why it is difficult to eliminate separation and reduce heat flow at the same time.

\backsection[Acknowledgements]{This work was supported by the National Key R \& D Program of China (Grant No. 2019YFA0405300), the China Postdoctoral Science Foundation (Grant Nos. 2021MD703969 and 2021MD703971), National Nature Science (China) Fund No. 12102449.}

\backsection[Declaration of interests]{The authors report no conflict of interest.}

%\bibliographystyle{jfm}
%\bibliography{jfm}
%Use of the above commands will create a bibliography using the .bib file. Shown below is a bibliography built from individual items.

\bibliographystyle{jfm}
\bibliography{jfm2esam}

\begin{thebibliography}{25}
\expandafter\ifx\csname natexlab\endcsname\relax\def\natexlab#1{#1}\fi
\def\au#1{#1} \def\ed#1{#1} \def\yr#1{#1}\def\at#1{#1}\def\jt#1{\textit{#1}}
  \def\bt#1{#1}\def\bvol#1{\textbf{#1}} \def\vol#1{#1} \def\pg#1{#1}
  \def\publ#1{#1}\def\arxiv#1{#1}\def\org#1{#1}\def\st#1{\textit{#1}}

\bibitem[{Babinsky} \& {Harvey}(2014)]{babinsky2014shock}
{\sc \au{{Babinsky}, Holger} \& \au{{Harvey}, John~K.}} \yr{2014} {\em Shock
  Wave-Boundary-Layer Interactions\/}.

\bibitem[{Biber} \& {Zumwalt}(1993)]{biber1993hysteresis}
{\sc \au{{Biber}, Kasim} \& \au{{Zumwalt}, Glen~W.}} \yr{1993}  \at{Hysteresis
  effects on wind tunnel measurements of a two-element airfoil}.  \jt{AIAA
  Journal}  \bvol{31}~(2),  \pg{326--330}.

\bibitem[{Chpoun} \& {Ben-Dor}(1995)]{chpoun1995numerical}
{\sc \au{{Chpoun}, A.} \& \au{{Ben-Dor}, G.}} \yr{1995}  \at{Numerical
  confirmation of the hysteresis phenomenon in the regular to the mach
  reflection transition in steady flows}.  \jt{Shock Waves}  \bvol{5}~(4),
  \pg{199--203}.

\bibitem[{Chpoun} {\em et~al.\/}(1995){Chpoun}, {Passerel}, {Li} \&
  {Ben-Dor}]{chpoun1995reconsideration}
{\sc \au{{Chpoun}, A.}, \au{{Passerel}, D.}, \au{{Li}, H.} \& \au{{Ben-Dor},
  G.}} \yr{1995}  \at{Reconsideration of oblique shock wave reflections in
  steady flows. part 1. experimental investigation}.  \jt{Journal of Fluid
  Mechanics}  \bvol{301}~(-1),  \pg{19--35}.

\bibitem[Fukagata {\em et~al.\/}(2002)Fukagata, Iwamoto \&
  Kasagi]{fukagata2002contribution}
{\sc \au{Fukagata, Koji}, \au{Iwamoto, Kaoru} \& \au{Kasagi, Nobuhide}}
  \yr{2002}  \at{Contribution of reynolds stress distribution to the skin
  friction in wall-bounded flows}.  \jt{Physics of Fluids}  \bvol{14}~(11),
  \pg{L73--L76}.

\bibitem[{Hornung} {\em et~al.\/}(1979){Hornung}, {Oertel} \&
  {Sandeman}]{hornung1979transition}
{\sc \au{{Hornung}, H.~G.}, \au{{Oertel}, H.} \& \au{{Sandeman}, R.~J.}}
  \yr{1979}  \at{Transition to mach reflexion of shock waves in steady and
  pseudosteady flow with and without relaxation}.  \jt{Journal of Fluid
  Mechanics}  \bvol{90}~(3),  \pg{541--560}.

\bibitem[Hu {\em et~al.\/}(2021)Hu, Wang, Zhou, Tang, Tang \&
  Yang]{hu2021existence}
{\sc \au{Hu, Yan-Chao}, \au{Wang, Gang}, \au{Zhou, Wen-Feng}, \au{Tang,
  Ming-Zhi}, \au{Tang, Zhi-Gong} \& \au{Yang, Yan-Guang}} \yr{2021} Existence
  of bistable states in curved compression ramp flows,  \arxiv{arXiv:
  2110.02033}.

\bibitem[{Hu} {\em et~al.\/}(2021){Hu}, {Zhou}, {Tang}, {Yang} \&
  {Qin}]{hu2021mechanism}
{\sc \au{{Hu}, Yan-Chao}, \au{{Zhou}, Wen-Feng}, \au{{Tang}, Zhi-Gong},
  \au{{Yang}, Yan-Guang} \& \au{{Qin}, Zhao-Hu}} \yr{2021}  \at{Mechanism of
  hysteresis in shock wave reflection.}  \jt{Physical Review E}
  \bvol{103}~(2),  \pg{23103--23103}.

\bibitem[{Hu} {\em et~al.\/}(2020){Hu}, {Zhou}, {Wang}, {Yang} \&
  {Tang}]{hu2020bistable}
{\sc \au{{Hu}, Yan-Chao}, \au{{Zhou}, Wen-Feng}, \au{{Wang}, Gang}, \au{{Yang},
  Yan-Guang} \& \au{{Tang}, Zhi-Gong}} \yr{2020}  \at{Bistable states and
  separation hysteresis in curved compression ramp flows}.  \jt{Physics of
  Fluids}  \bvol{32}~(11),  \pg{113601}.

\bibitem[{Ivanov} {\em et~al.\/}(2001){Ivanov}, {Ben-Dor}, {Elperin},
  {Kudryavtsev} \& {Khotyanovsky}]{ivanov2001flow}
{\sc \au{{Ivanov}, M.~S.}, \au{{Ben-Dor}, G.}, \au{{Elperin}, T.},
  \au{{Kudryavtsev}, A.~N.} \& \au{{Khotyanovsky}, D.~V.}} \yr{2001}
  \at{Flow-mach-number-variation-induced hysteresis in steady shock wave
  reflections}.  \jt{AIAA Journal}  \bvol{39}~(5),  \pg{972--974}.

\bibitem[Li {\em et~al.\/}(2010)Li, Fu, Ma \& Liang]{li2010direct}
{\sc \au{Li, XinLiang}, \au{Fu, DeXun}, \au{Ma, YanWen} \& \au{Liang, Xian}}
  \yr{2010}  \at{Direct numerical simulation of shock/turbulent boundary layer
  interaction in a supersonic compression ramp}.  \jt{Science China Physics,
  Mechanics and Astronomy}  \bvol{53}~(9),  \pg{1651--1658}.

\bibitem[McCroskey(1982)]{mccroskey1982unsteady}
{\sc \au{McCroskey, William~J}} \yr{1982}  \at{Unsteady airfoils}.  \jt{Annual
  review of fluid mechanics}  \bvol{14}~(1),  \pg{285--311}.

\bibitem[{Mittal} \& {Saxena}(2000)]{mittal2000prediction}
{\sc \au{{Mittal}, S.} \& \au{{Saxena}, P.}} \yr{2000}  \at{Prediction of
  hysteresis associated with the static stall of an airfoil}.  \jt{AIAA
  Journal}  \bvol{38}~(5),  \pg{933--935}.

\bibitem[{Mueller}(1985)]{mueller1985the}
{\sc \au{{Mueller}, Thomas~J.}} \yr{1985}  \at{The influence of laminar
  separation and transition on low reynolds number airfoil hysteresis}.
  \jt{Journal of Aircraft}  \bvol{22}~(9),  \pg{763--770}.

\bibitem[Renard \& Deck(2016)]{renard2016theoretical}
{\sc \au{Renard, Nicolas} \& \au{Deck, S{\'e}bastien}} \yr{2016}  \at{A
  theoretical decomposition of mean skin friction generation into physical
  phenomena across the boundary layer}.  \jt{Journal of Fluid Mechanics}
  \bvol{790},  \pg{339--367}.

\bibitem[{Simeonides} \& {Haase}(1995)]{simeonides1995experimental}
{\sc \au{{Simeonides}, G.} \& \au{{Haase}, W.}} \yr{1995}  \at{Experimental and
  computational investigations of hypersonic flow about compression ramps}.
  \jt{Journal of Fluid Mechanics}  \bvol{283}~(-1),  \pg{17--42}.

\bibitem[{Simeonides} {\em et~al.\/}(1994){Simeonides}, {Haase} \&
  {Manna}]{simeonides1994experimental}
{\sc \au{{Simeonides}, G.}, \au{{Haase}, W.} \& \au{{Manna}, M.}} \yr{1994}
  \at{Experimental, analytical, and computational methods applied to hypersonic
  compression ramp flows}.  \jt{AIAA Journal}  \bvol{32}~(2),  \pg{301--310}.

\bibitem[Sun {\em et~al.\/}(2021)Sun, Guo, Yuan, Zhang, Li \&
  Liu]{sun2021decomposition}
{\sc \au{Sun, Dong}, \au{Guo, Qilong}, \au{Yuan, Xianxu}, \au{Zhang, Haoyuan},
  \au{Li, Chen} \& \au{Liu, Pengxin}} \yr{2021}  \at{A decomposition formula
  for the wall heat flux of a compressible boundary layer}.  \jt{Advances in
  Aerodynamics}  \bvol{3}~(1),  \pg{1--13}.

\bibitem[Tang {\em et~al.\/}(2021)Tang, Wang, Xie, Zhou, Hu \&
  Yang]{tang2021aerothermodynamic}
{\sc \au{Tang, Ming-Zhi}, \au{Wang, Gang}, \au{Xie, Zhu-Xuan}, \au{Zhou,
  Wen-Feng}, \au{Hu, Yan-Chao} \& \au{Yang, Yan-Guang}} \yr{2021}
  \at{Aerothermodynamic characteristics of hypersonic curved compression ramp
  flows with bistable states}.  \jt{Physics of Fluids}  \bvol{33}~(12),
  \pg{126106}.

\bibitem[Vuillon {\em et~al.\/}(1995)Vuillon, Zeitoun \&
  Ben-Dor]{vuillon1995reconsideration}
{\sc \au{Vuillon, J}, \au{Zeitoun, D} \& \au{Ben-Dor, G}} \yr{1995}
  \at{Reconsideration of oblique shock wave reflections in steady flows. part
  2. numerical investigation}.  \jt{Journal of Fluid Mechanics}  \bvol{301},
  \pg{37--50}.

\bibitem[{Yang} {\em et~al.\/}(2008){Yang}, {Igarashi}, {Martin} \&
  {Hu}]{yang2008an}
{\sc \au{{Yang}, Zifeng}, \au{{Igarashi}, Hirofumi}, \au{{Martin}, Mathew} \&
  \au{{Hu}, Hui}} \yr{2008} An experimental investigation on aerodynamic
  hysteresis of a low-reynolds number airfoil.  \bt{In {\em 46th AIAA Aerospace
  Sciences Meeting and Exhibit\/}}.

\bibitem[{Zhang}(2020)]{zhang2020hypersonic}
{\sc \au{{Zhang}, Kunyuan}} \yr{2020} {\em Hypersonic Curved Compression Inlet
  and Its Inverse Design\/}.

\bibitem[Zhang {\em et~al.\/}(2022)Zhang, Song \& Xia]{zhang2022exact}
{\sc \au{Zhang, Peng}, \au{Song, Yubin} \& \au{Xia, Zhenhua}} \yr{2022}
  \at{Exact mathematical formulas for wall-heat flux in compressible turbulent
  channel flows}.  \jt{Acta Mechanica Sinica}  \bvol{38}~(1),  \pg{1--10}.

\bibitem[Zhang \& Xia(2020)]{zhang2020contribution}
{\sc \au{Zhang, Peng} \& \au{Xia, Zhenhua}} \yr{2020}  \at{Contribution of
  viscous stress work to wall heat flux in compressible turbulent channel
  flows}.  \jt{Physical Review E}  \bvol{102}~(4),  \pg{043107}.

\bibitem[{Zhou} {\em et~al.\/}(2021){Zhou}, {Hu}, {Tang}, {Wang}, {Fang} \&
  {Yang}]{zhou2021mechanism}
{\sc \au{{Zhou}, Wen-Feng}, \au{{Hu}, Yan-Chao}, \au{{Tang}, Ming-Zhi},
  \au{{Wang}, Gang}, \au{{Fang}, Ming} \& \au{{Yang}, Yan-Guang}} \yr{2021}
  \at{Mechanism of separation hysteresis in curved compression ramp.}
  \jt{Physics of Fluids}  \bvol{33}~(10),  \pg{106108}.

\end{thebibliography}

%\begin{thebibliography}{99}
%
%\expandafter\ifx\csname natexlab\endcsname\relax
%\def\natexlab#1{#1}\fi
%\expandafter\ifx\csname selectlanguage\endcsname\relax
%\def\selectlanguage#1{\relax}\fi
%
%\bibitem[Batchelor (1971)]{Batchelor59}
%{\sc Batchelor, G.K.} 1971 {Small-scale variation of convected quantities like temperature in turbulent fluid part1, general discussion and the case of small conductivity}, {\it J. Fluid Mech.}, {\bf 5}, pp. 3-113-133.
%
%
%\end{thebibliography}

% End of file `jfm2esam.bib'.

\end{document}